\documentclass[
    10pt,
    onecolumn,
    secnumarabic,
    superscriptaddress,
    nobibnotes,
    nofootinbib,
    showkeys,
    aps,
    prd
]{revtex4-2}

\usepackage[T1]{fontenc}       % Better font encoding
\usepackage{amsmath, amssymb}  % Math symbols and environments
\usepackage{amsfonts}          % Extra math fonts
\usepackage{graphicx}          % Figures and graphics
\usepackage{bm}                % Bold math symbols
\usepackage{xcolor}            % Color support
\usepackage{comment}           % Commenting out blocks
\usepackage{hyperref}          % Hyperlinks
\usepackage[justification=justified,singlelinecheck=false,labelsep=period]{caption}
\usepackage{enumitem}
\usepackage{subcaption}
\providecommand{\U}[1]{\protect\rule{.1in}{.1in}}

\usepackage{tocloft} % For customizing table of contents, figures, and tables

\usepackage{booktabs}
\usepackage{multirow}
\usepackage{siunitx}
\usepackage{array}

\begin{document}

\title{Late-time Background Constraints on Linear and Non-linear  Interacting Dark Energy after DESI DR2}
\author{David Figueruelo}
\email{david.figueruelo@ehu.eus}
\affiliation{Department of Theoretical Physics, University of the Basque Country UPV/EHU, 48080 Bilbao, Spain}
\affiliation{Institute of Theoretical Astrophysics, University of Oslo, N-0315 Oslo, Norway}

\author{Marcel van der Westhuizen}
\email{marcelvdw007@gmail.com }
\affiliation{Centre for Space Research, North-West University, Potchefstroom 2520, South Africa}

\author{Amare Abebe}
\email{amare.abebe@nithecs.ac.za}
\affiliation{Centre for Space Research, North-West University, Potchefstroom 2520, South
Africa}
\affiliation{National Institute for Theoretical and Computational Sciences (NITheCS),
South Africa}

\author{Eleonora Di Valentino}
\email{e.divalentino@sheffield.ac.uk}
\affiliation{School of Mathematical and Physical Sciences, University of Sheffield, Hounsfield Road, Sheffield S3 7RH, United Kingdom}

\begin{abstract}
In this study, we present observational constraints on a class of phenomenological interacting dark energy (IDE) models that admit analytical solutions for the Hubble parameter $H(z)$. We consider a set of five linear and three non-linear IDE scenarios, encompassing both interactions proportional to the dark matter and/or dark energy densities, as well as non-linear combinations of the two. For all eight IDE models, we find a better fit than $\Lambda$CDM from a $\Delta\chi^2$ analysis for both combinations of datasets considered. When using the Akaike Information Criterion ($\Delta$AIC), we find a similarly improved fit in all cases, except for one dataset combination in $Q=3H\delta\rho_{\rm de}$. Our analysis also shows a preference for sign-switching interactions, with energy transfer from dark energy to dark matter at low redshift, reversing direction at higher redshift. These results should be interpreted with caution, as the latter direction of energy transfer is accompanied by negative dark energy densities in the past, which may be unphysical. Models that do not allow sign-changing behaviour instead show a preference for energy flow from dark matter to dark energy, and hence negative dark energy densities. The only exceptions are $Q=3H\delta\rho_{\rm de}$ and $Q=3H\delta\left(\tfrac{\rho_{\rm de}^2}{\rho_{\rm dm}+\rho_{\rm de}}\right)$, which exhibit energy flow in the opposite direction. Furthermore, for all interactions considered, we find a phantom-divide crossing in the effective dark energy equation of state $w^{\rm eff}_{\rm de}$, with the dark energy density decreasing ($w^{\rm eff}_{\rm de}>-1$) at present and at low redshift, while increasing ($w^{\rm eff}_{\rm de}<-1$) in the past at high redshift. These results highlight the promising, but problematic, nature of dark sector interactions, as well as the need to extend the analysis using early-time physics and datasets.

\end{abstract}
\maketitle

%\clearpage \newpage
\section{Introduction}

Cosmology has entered a new era driven by an unprecedented wealth of observational data collected in recent years. This change has reshaped the way cosmology itself develops as a science, since any new proposal must now be evaluated against the plethora of available datasets. Large galaxy surveys such as DESI~\cite{DESI:2025zgx,DESI:2025zpo} or DES~\cite{DES:2025key}, extraordinarily precise measurements of the Cosmic Microwave Background from Planck~\cite{Planck:2018vyg}, and Supernovae light curve observations~\cite{Brout:2022vxf,DES:2025sig}, among many others, have elevated cosmology to a discipline built upon exquisite and precise observations.
The diversity of these measurements has provided us with a map of the Universe across different epochs and scales, while allowing us to test and rule out alternative models with a level of precision unimaginable just a few decades ago. These data have established the $\Lambda$ Cold Dark Matter model ($\Lambda$CDM) as the standard cosmological model, yet they have also highlighted its most profound mystery: the required dominance of a dark sector. This dark sector, however, comes with its own theoretical challenges~\cite{Martin:2012bt,Weinberg:1988cp,Zlatev:1998tr,Velten:2014nra}. At the frontier of modern cosmology lies this dark sector, a realm whose elusive nature continues to challenge our deepest attempts to unveil the fundamental laws of the Universe.

In addition to this mystery, we have reached a point where such an abundance of precise data has led to the emergence of tensions in cosmological measurements~\cite{Verde:2019ivm,DiValentino:2020zio,DiValentino:2021izs,Perivolaropoulos:2021jda,Schoneberg:2021qvd,Shah:2021onj,Abdalla:2022yfr,DiValentino:2022fjm,Kamionkowski:2022pkx,Giare:2023xoc,Hu:2023jqc,Verde:2023lmm,DiValentino:2024yew,CosmoVerse:2025txj,Ong:2025cwv}, with the Hubble $H_0$ tension being of extraordinary concern and the $S_8$ tension of comparatively lesser significance. One possible explanation for these discrepancies is their origin in observational systematics or miscalibrations; however, as data quality improves and additional datasets become available, this possibility is increasingly disfavoured~\cite{ACT:2025fju,SPT-3G:2025bzu,H0DN:2025lyy,Porredon:2025xxv}.
If systematics are not the cause, these tensions are rendered more pressing by the \emph{trend} they exhibit: values inferred from early-Universe experiments, where $\Lambda$CDM is deeply embedded in the analysis, versus values obtained from late-time measurements that rely far less on the $\Lambda$CDM framework, rather than by the nominal statistical significance (in $\sigma$) separating the two measurements. This point is not inconsequential but rather pivotal if we are to truly understand the current situation. It therefore raises a central question: to what extent are these tensions the result of a potential incompleteness of the $\Lambda$CDM model?
In addition to these long-standing cosmological tensions, the recent DESI data~\cite{DESI:2025zgx,DESI:2025zpo} have introduced a possible new discrepancy by suggesting a preference for dynamical dark energy, although this issue remains under active debate. Within this context, the study of alternatives to the reference $\Lambda$CDM model is well motivated: first, as a means of better understanding the standard model itself, and second, as a pathway toward potentially improved descriptions of our Universe.

The landscape of modifications to the $\Lambda$CDM model is as vast as our imagination allows, but it can be broadly summarised as follows: modifications or reformulations of gravity~\cite{DiValentino:2015bja,Zumalacarregui:2020cjh,Odintsov:2020qzd,Adi:2020qqf,DeFelice:2020cpt,Pogosian:2021mcs,CANTATA:2021asi,Schiavone:2022wvq,Ishak:2024jhs,Specogna:2023nkq,Specogna:2024euz,AtacamaCosmologyTelescope:2025nti,Giare:2025ath,Tiwari:2023jle,Hogas:2023pjz,Wen:2023wes,Pitrou:2023swx,Montani:2024pou,Dwivedi:2024okk,Akarsu:2024qsi,Akarsu:2024nas,Hogas:2025ahb}, dynamical dark energy~\cite{DESI:2025fii,DESI:2024mwx,DESI:2025zgx,Cortes:2024lgw,Shlivko:2024llw,Luongo:2024fww,Gialamas:2024lyw,Dinda:2024kjf,Najafi:2024qzm,Wang:2024dka,Ye:2024ywg,Tada:2024znt,Carloni:2024zpl,Chan-GyungPark:2024mlx,DESI:2024kob,Bhattacharya:2024hep,Ramadan:2024kmn,Pourojaghi:2024tmw,Giare:2024gpk,Reboucas:2024smm,Giare:2024ocw,Chan-GyungPark:2024brx,Li:2024qus,Jiang:2024xnu,RoyChoudhury:2024wri,Li:2025cxn,Wolf:2025jlc,Shajib:2025tpd,Giare:2025pzu,Chaussidon:2025npr,Kessler:2025kju,Pang:2025lvh,Roy:2024kni,RoyChoudhury:2025dhe,Paliathanasis:2025cuc,Scherer:2025esj,Giare:2024oil,Liu:2025mub,Teixeira:2025czm,Santos:2025wiv,Specogna:2025guo,Sabogal:2025jbo,Cheng:2025lod,Herold:2025hkb,Cheng:2025hug,Ozulker:2025ehg,Lee:2025pzo,Ormondroyd:2025iaf,Silva:2025twg,Ishak:2025cay,Fazzari:2025lzd,Smith:2025icl,Zhang:2025lam,Cheng:2025yue}, the introduction of new species~\cite{Tulin:2017ara,Buen-Abad:2015ova,DiValentino:2017oaw,Anchordoqui:2022gmw,Pan:2023frx,Allali:2024anb,Co:2024oek,Aboubrahim:2024spa}, non-gravitational interactions involving the dark sector~\cite{Kumar:2016zpg,Murgia:2016ccp,Kumar:2017dnp,DiValentino:2017iww,Kumar:2021eev,Gao:2021xnk,Pan:2023mie,Benisty:2024lmj,Yang:2020uga,Forconi:2023hsj,Pourtsidou:2016ico,DiValentino:2020vnx,DiValentino:2020leo,Nunes:2021zzi,Yang:2018uae,vonMarttens:2019ixw,Lucca:2020zjb,Wang:2024vmw,Gao:2022ahg,Zhai:2023yny,Bernui:2023byc,Becker:2020hzj,Hoerning:2023hks,Giare:2024ytc,Escamilla:2023shf,vanderWesthuizen:2023hcl,Silva:2024ift,DiValentino:2019ffd,Li:2024qso,Pooya:2024wsq,Halder:2024uao,Castello:2023zjr,Yao:2023jau,Mishra:2023ueo,Nunes:2016dlj,Silva:2025hxw,Yang:2025uyv,vanderWesthuizen:2025vcb,vanderWesthuizen:2025mnw,vanderWesthuizen:2025rip}, modifications at recombination~\cite{Lee:2022gzh,Greene:2023cro,Greene:2024qis,Baryakhtar:2024rky,Seto:2024cgo,Lynch:2024hzh,Toda:2024ncp,Schoneberg:2024ynd,Smith:2025uaq}, and Early Dark Energy or other early-time solutions~\cite{Poulin:2018cxd,Smith:2019ihp,Niedermann:2019olb,Krishnan:2020obg,Schoneberg:2021qvd,Ye:2021iwa,Poulin:2021bjr,Niedermann:2021vgd,deSouza:2023sqp,Poulin:2023lkg,Cruz:2023lmn,Niedermann:2023ssr,Vagnozzi:2023nrq,Efstathiou:2023fbn,Cervantes-Cota:2023wet,Garny:2024ums,Giare:2024akf,Giare:2024syw,Poulin:2024ken,Pedrotti:2024kpn,Kochappan:2024jyf}.
Although the idea of dark sector interactions was originally proposed to address the coincidence problem, such interactions have more recently been put forward as potential explanations for the aforementioned cosmological tensions.

In this study, we focus on the specific case of a non-gravitational energy transfer within the dark sector, broadly known as Interacting Dark Energy (IDE) models. Here, we restrict our analysis to the background evolution, leaving the study of perturbations for future work. This paper should therefore be understood as a continuation of Refs.~\cite{vanderWesthuizen:2025vcb,vanderWesthuizen:2025mnw,vanderWesthuizen:2025rip}. As previously mentioned, these models were introduced shortly after the discovery of the accelerated expansion of the Universe to address the coincidence problem~\cite{Amendola:1999er,Zimdahl:2001ar,Chimento:2003iea,Farrar:2003uw,Wang:2004cp,Olivares:2006jr,Sadjadi:2006qp,Quartin:2008px,delCampo:2008jx,Caldera-Cabral:2008yyo,He:2010im,delCampo:2015vha}.
The coincidence problem arises from the different dilution rates of the dark components: the cosmological constant does not dilute ($\rho_{\Lambda}=\mathrm{constant}$), while dark matter dilutes proportionally to the expansion of the Universe ($\rho_{\rm dm}\propto a^{-3}$). Despite this striking difference in dilution rates, observers in our Universe appear to live at a particular epoch in which both components are of the same order of magnitude. The distant past and future correspond to entirely different scenarios, which naturally raises the question: why now?
Following their initial introduction, IDE models experienced a resurgence in popularity driven by the persistent nature of the aforementioned cosmological tensions and the search for possible extensions to the $\Lambda$CDM model. In particular, the potential of IDE models to address the $H_0$ tension has been explored extensively~\cite{Benisty:2024lmj,Forconi:2023hsj,Nunes:2021zzi,Zhai:2023yny,Bernui:2023byc,Hoerning:2023hks,Giare:2024ytc,Escamilla:2023shf,Li:2024qso,Pooya:2024wsq,Halder:2024uao,Castello:2023zjr,Yao:2023jau,Mishra:2023ueo,Silva:2025hxw,DiValentino:2021izs,Califano:2022syd,Pan:2023mie,Liu:2023wew,Sabogal:2025mkp,Yang:2025uyv,Vagnozzi:2023nrq}, while the impact of interacting dark sectors on the $S_8$ tension has been studied in~\cite{Kumar:2017dnp,Kumar:2019wfs,DiValentino:2019ffd,Anchordoqui:2021gji,Kumar:2021eev,Gariazzo:2021qtg,Lucca:2021dxo,Sabogal:2024yha,Liu:2023kce,Yang:2025ume,Liu:2025vda}.
More recently, claims of a possible dynamical nature of dark energy emerging from the DESI DR2\footnote{Similar results were previously reported in the DESI DR1 data release~\cite{DESI:2024mwx}.}
data release~\cite{DESI:2025zgx,DESI:2025fii,DESI:2025qqy,DESI:2025wyn} have prompted renewed interest in models capable of simultaneously providing a dynamical description of dark energy and allowing for a phantom crossing. This has led to further investigations of IDE models as viable candidates to address key problems in cosmology~\cite{Guedezounme:2025wav,Silva:2025hxw,Pan:2025qwy,Shah:2025ayl,Lee:2025pzo,vanderWesthuizen:2025iam,Giare:2024smz,Zhu:2025lrk,Gonzalez-Espinoza:2025vrc,MV:2025yjt,Lyu:2025nsd,Wu:2025vrl,Li:2025ula}. Other approaches aimed at explaining the DESI DR2 results can be found in~\cite{Wang:2025znm,Yao:2025wlx,Bhattacharya:2024hep,Bhattacharya:2024kxp}.
%\EDV{References are never too much. It is a good way to advertise a paper. :) }

Our study builds upon these investigations by specifically constraining a set of phenomenological IDE models using the same BAO data from DESI DR2, together with cosmological data from Pantheon+, Cosmic Clocks, and BBN. The models considered here have recently been studied in~\cite{vanderWesthuizen:2025vcb,vanderWesthuizen:2025rip,vanderWesthuizen:2025mnw}, where new analytical solutions for the Hubble function were derived, along with conditions describing the occurrence of negative energy densities and future singularities.
This work can therefore be regarded as the first set of observational constraints on these models using the most up-to-date data following these theoretical developments, enabling us to draw new conclusions regarding a possible preference in cosmological data for negative energy densities within the framework of IDE models.

This paper is structured as follows. In Section~\ref{sec:IDE}, we introduce the interacting dark energy models considered in this work and describe their theoretical framework. In Section~\ref{sec:method}, we present the datasets and methodology used in our analysis, together with the resulting observational constraints. Finally, we summarize our main conclusions and outline future prospects in Section~\ref{sec:conclusions}.

\section{Interacting dark energy models} \label{sec:IDE}

In this section, we introduce the models considered in this work and present the equations that govern their dynamics. For detailed discussions of the models and the derivation of the equations presented here, we refer the reader to Refs.~\cite{vanderWesthuizen:2025vcb,vanderWesthuizen:2025mnw,vanderWesthuizen:2025rip}, where the theoretical analyses are carried out in full.

Throughout this work, we assume that the different components of the Universe can be described as perfect fluids. Accordingly, we associate with each component an energy density $\rho_i$, a pressure $p_i$, and a four-velocity $u_i^\mu$. We also assign to each component a stress-energy tensor $T_i^{\mu\nu}$, which is covariantly conserved for each individual, non-interacting component, $\nabla_\mu T_i^{\mu\nu}=0$.
However, for the interacting dark energy and dark matter components considered here, the conservation equations take the form
\begin{equation}\label{conservation}
\nabla_\mu T^{\mu\nu}_{\rm de} = - \nabla_\mu T^{\mu\nu}_{\rm dm} = Q^\nu \, .
\end{equation}
Here, $Q^\nu$ denotes the interaction four-vector and represents the flow of energy between the two dark sectors. Since we are only interested in the background evolution, we do not require the full four-vector $Q^\nu$, but only its background zeroth component, which we denote simply by $Q$.
In this paper, we consider a variety of interacting models in which the coupling kernel $Q$ is proportional, either linearly or non-linearly, to the dark sector densities. The dynamics of these models are governed by the modified conservation equation~\eqref{conservation}, which at the background level leads to the following set of conservation equations:
\begin{equation}\label{eq:conservation}
\dot{\rho}_{\rm dm} + 3H\rho_{\rm dm} = Q \,, \qquad
\dot{\rho}_{\rm de} + 3H\rho_{\rm de}(1+w) = -Q \, .
\end{equation}
Here, the overdot denotes differentiation with respect to cosmic time, $\rho_{\rm dm}$ and $\rho_{\rm de}$ are the energy densities of dark matter and dark energy, respectively, $H$ is the Hubble function, and $w$ is the dark energy equation-of-state parameter. Dark matter is assumed to be pressureless.
From these equations, it is clear that $Q>0$ corresponds to an energy transfer from dark energy to dark matter, while $Q<0$ describes energy flow from dark matter to dark energy. The direction of this energy exchange can significantly affect both the background dynamics and the expansion history of the Universe. 
Assuming that all other parameters are kept fixed to their $\Lambda$CDM values and that only the interaction is introduced, we expect the following deviations from standard cosmology:
\begin{enumerate}
    \item \textbf{Energy flow from dark energy to dark matter:} 
    This scenario leads to less dark matter and more dark energy in the past, thereby \textit{alleviating the coincidence problem}. The suppression of the dark matter density results in a lower value of $H_0$ at late times, thus \textit{worsening the $H_0$ tension}. In addition, the matter power spectrum is suppressed with respect to late-time probes, which \textit{alleviates the $S_8$ tension}~\cite{Lucca:2021dxo}. Energy flow in this direction also \textit{increases the age of the Universe} and tends to \textit{avoid both negative energy densities and future big rip singularities}~\cite{vanderWesthuizen:2025vcb,vanderWesthuizen:2025mnw}, while remaining consistent with the laws of thermodynamics~\cite{Pavon:2007gt}.
    
    \item \textbf{Energy flow from dark matter to dark energy:} 
    Energy transfer in this direction generally produces the opposite effects to those described above. Nevertheless, it is often considered attractive due to its potential to alleviate the Hubble tension, as demonstrated in~\cite{Salvatelli:2013wra,Kumar:2016zpg, Caprini:2016qxs, Murgia:2016ccp, Zheng:2017asg, Kumar:2017dnp, DiValentino:2017iww,Kumar:2021eev,Gao:2021xnk,Pan:2023mie,Benisty:2024lmj,Yang:2020uga,Forconi:2023hsj,Pourtsidou:2016ico,DiValentino:2020vnx,DiValentino:2020leo,Nunes:2021zzi,Yang:2018uae,Zhang:2018mlj,vonMarttens:2019ixw,Lucca:2020zjb,Xiao:2021nmk,Wang:2024vmw,Gao:2022ahg,Zhai:2023yny,Joseph:2022khn,Bernui:2023byc,Becker:2020hzj,Hoerning:2023hks,Giare:2024ytc,Mukhopadhyay:2020bml,Escamilla:2023shf,vanderWesthuizen:2023hcl,Silva:2024ift,DiValentino:2019ffd,Zhao:2022ycr,Li:2024qso,Pooya:2024wsq,Halder:2024uao,Castello:2023zjr,Yao:2023jau,Mishra:2023ueo,Nunes:2016dlj,Silva:2025hxw,Yang:2025uyv,vanderWesthuizen:2025rip}.
\end{enumerate}
It should be emphasized that many of the consequences outlined above hold only if all cosmological parameters are kept fixed. Once other parameters, such as $\Omega_{\rm dm,0}$, $H_0$, and $w$, are allowed to vary freely, different conclusions may be reached depending on their inferred posterior distributions.
Another important consideration is that some interactions exhibit a sign-switching behaviour, which can lead to combinations of the consequences described above.

\subsection{Linear models}
\label{subsec:model_linear}

In this analysis, we consider five interaction kernels proportional in a linear manner to either the dark matter or dark energy density, or a combination of both. The background dynamics of these interactions were recently studied in~\cite{vanderWesthuizen:2025vcb}. Specifically, we consider the general interaction given by the kernel $Q = 3H(\delta_{\rm dm}\rho_{\rm dm} + \delta_{\rm de}\rho_{\rm de})$ and four of its special cases.

\subsubsection*{Linear IDE model 1: $Q=3 H (\delta_{\rm{dm}} \rho_{\rm{dm}} + \delta_{\rm{de}}  \rho_{\rm{de}})$}

For a flat FLRW universe containing radiation, baryons, dark matter, and dark energy, this interaction has the normalized Hubble function found in Ref.~\cite{vanderWesthuizen:2025rip}:
\begin{gather}
\begin{split} \label{hz_Q_linear}
E(z) = \Bigg\{ -& \frac{1}{2 \Delta} 
\Bigl[\Omega_{\text{(de,0)}}\Bigl(\delta_{\text{dm}}-\delta_{\text{de}}+w-\Delta\Bigr)
+\Omega_{\text{(dm,0)}}\Bigl(\delta_{\text{dm}}-\delta_{\text{de}}-w-\Delta\Bigr)\Bigr] 
(1+z)^{-\tfrac{3}{2}\Bigl(\delta_{\text{dm}}-\delta_{\text{de}}-w-2+\Delta\Bigr)} \\[1mm]
+& \frac{1}{2 \Delta} 
\Bigl[\Omega_{\text{(de,0)}}\Bigl(\delta_{\text{dm}}-\delta_{\text{de}}+w+\Delta\Bigr)
+\Omega_{\text{(dm,0)}}\Bigl(\delta_{\text{dm}}-\delta_{\text{de}}-w+\Delta\Bigr)\Bigr] 
(1+z)^{-\tfrac{3}{2}\Bigl(\delta_{\text{dm}}-\delta_{\text{de}}-w-2-\Delta\Bigr)} \\[1mm]
+& \Omega_{\text{bm,0}}(1+z)^3 + \Omega_{\text{r,0}}(1+z)^4 
\Bigg\}^{\tfrac{1}{2}},
\end{split}   
\end{gather}
where $\Omega_{\text{(dm,0)}}$ and $\Omega_{\text{(de,0)}}$ are the present ($z=0$) density parameters for dark matter and dark energy, $\delta_{\text{dm}}$ and $\delta_{\text{de}}$ are dimensionless coupling constants that determine how strongly the interaction is coupled to dark matter and dark energy, respectively, while $w$ is the dark energy equation of state. We also assume that both baryonic matter and dark matter have no pressure. Notably, the determinant $\Delta$ appears and is given by:
\begin{gather}
\begin{split} \label{eq:determinant_Q_linear}
\Delta = \sqrt{(\delta_{\text{dm}}+\delta_{\text{de}}+w)^2 - 4\delta_{\text{de}}\delta_{\text{dm}}}\,. 
\end{split}   
\end{gather} 
The sign of $\delta_{\rm dm}$ determines the initial direction of energy transfer, while $\delta_{\rm de}$ determines the late-time energy transfer. In both cases, a positive coupling implies energy transfer from dark energy to dark matter, and vice versa for a negative coupling. This implies that if $\delta_{\rm dm}$ and $\delta_{\rm de}$ have opposite signs, the direction of energy transfer will change at some point during the cosmological evolution. Furthermore, $\delta_{\rm dm}<0$ leads to negative dark energy in the past, while $\delta_{\rm de}<0$ predicts negative dark matter in the future, as shown in~\cite{vanderWesthuizen:2025vcb}.

\subsubsection*{Linear IDE model 2: $Q=3H\delta(\rho_{\rm{dm}}+\rho_{\rm{de}})$}

The Hubble function for this model is given by \eqref{hz_Q_linear} when we set $\delta_{\text{dm}}=\delta_{\text{de}}=\delta$. This model exhibits negative dark energy densities in the past and negative dark matter densities in the future if energy flows from dark matter to dark energy (given by $\delta<0$), while all energy densities remain positive if there is a small interaction from dark energy to dark matter (given by $\delta>0$), provided the conditions in Table~\ref{tab:Com_PEC} are satisfied.

Previous observational constraints for this model may be found in~\cite{Costa:2013sva,Costa:2016tpb,An:2017kqu,An:2017crg,Bachega:2019fki,Li:2019loh,Halder:2021jiv,Mukhopadhyay:2020bml,Xiao:2021nmk,Aljaf:2020eqh,Califano:2024xzt,Wu:2025vrl}.

\subsubsection*{Linear IDE model 3: $Q=3H \delta (\rho_{\rm{dm}}-\rho_{\rm{de}})$}

The Hubble function for this model is given by \eqref{hz_Q_linear} when we set $\delta_{\text{dm}}=\delta$ and $\delta_{\text{de}}=-\delta$. This is a sign-switching interaction, with the sign of $\delta$ determining the initial direction of energy transfer. If $\delta<0$, the energy initially flows from dark matter to dark energy and reverses at some point during the evolution. This behaviour is also accompanied by negative dark energy densities in the past. In contrast, if $\delta>0$, the initial energy flow is from dark energy to dark matter, which later changes to energy transfer from dark matter to dark energy. This scenario further predicts that dark matter becomes negative in the future.

Previous observational constraints for this model may be found in~\cite{Aljaf:2020eqh,Pan:2019jqh,Pan:2023mie}.

\subsubsection*{Linear IDE model 4: $Q=3H\delta\rho_{\rm{dm}}$}

The Hubble function for this model is given by \eqref{hz_Q_linear} when we set $\delta_{\text{dm}}=\delta$ and $\delta_{\text{de}}=0$. If energy flows from dark matter to dark energy (given by $\delta<0$), this interaction leads to negative dark energy densities in the past, while all energy densities remain positive for a small interaction from dark energy to dark matter (given by $\delta>0$).

Previous observational constraints for this model may be found in~\cite{Kumar:2016zpg,Li:2020gtk,Guo:2021rrz,DiValentino:2021izs,Wang:2021kxc,Costa:2013sva,Costa:2016tpb,Santos:2017bqm,An:2017kqu,An:2017crg,Grandon:2018uoe,vonMarttens:2018iav,Bachega:2019fki,Aljaf:2020eqh,Califano:2024xzt,Li:2024qso,Benisty:2024lmj,Yan:2025iga,Nagpal:2025hnl,Wu:2025vrl}.

\subsubsection*{Linear IDE model 5: $Q=3H\delta\rho_{\rm{de}}$}

The Hubble function for this model is given by \eqref{hz_Q_linear} when we set $\delta_{\text{de}}=\delta$ and $\delta_{\text{dm}}=0$. If energy flows from dark matter to dark energy (given by $\delta<0$), this interaction predicts negative dark matter densities in the future, while all energy densities remain positive for a small interaction from dark energy to dark matter (given by $\delta>0$).

Previous observational constraints for this model may be found in~\cite{Salvatelli:2013wra,DiValentino:2017iww,Kumar:2017dnp,Kumar:2019wfs,Li:2020gtk,Yang:2019uog,Kumar:2021eev,Lucca:2020zjb,DiValentino:2019jae,DiValentino:2019ffd,DiValentino:2021izs,Lucca:2021dxo,Anchordoqui:2021gji,Wang:2021kxc,Yang:2022csz,Bernui:2023byc,Califano:2022syd,Pan:2023mie,Sabogal:2024yha,Giare:2024smz,Sabogal:2025mkp,Clemson:2011an,Costa:2013sva,Costa:2016tpb,An:2017kqu,Santos:2017bqm,An:2017crg,Grandon:2018uoe,vonMarttens:2018iav,Aljaf:2020eqh,Nunes:2022bhn,Zhai:2023yny,Li:2024qso,Yan:2025iga,Wu:2025vrl}.

\subsection{Non-linear models}
 \label{subsec:model_non_linear}
 
In this section, we present three models in which the interaction depends on some product of the dark matter and dark energy densities, whose background dynamics were recently studied in depth in~\cite{vanderWesthuizen:2025mnw}. The three models discussed below can be seen as special cases of $Q=3H\delta \left( \frac{\rho_{\text{dm}}^\alpha \rho_{\text{de}}^\beta }{\rho_{\text{dm}}+\rho_{\text{de}}} \right)$, with $(\alpha,\beta)=(1,1)$, $(\alpha,\beta)=(2,0)$, and $(\alpha,\beta)=(0,2)$ corresponding to non-linear IDE models~1, 2, and~3, respectively.

\subsubsection*{Non-linear IDE model 1: $Q=3H\delta \left( \frac{\rho_{\text{dm}} \rho_{\text{de}} }{\rho_{\text{dm}}+\rho_{\text{de}}} \right)$}
 \label{subsec:model_non_linear_1}

The normalized Hubble function for this model is found in~\cite{vanderWesthuizen:2025rip, Arevalo:2011hh} and is given by:
\begin{gather}
\begin{split} \label{hz_Q_dmde}
E(z) = \Bigg\{ & \Bigl[\Omega_{\text{(dm,0)}}(1+z)^{3(1-\delta)} 
+ \Omega_{\text{(de,0)}}(1+z)^{3(1+w)}\Bigr] 
\left[\frac{1+\left(\tfrac{\Omega_{\text{(dm,0)}}}{\Omega_{\text{(de,0)}}}\right)(1+z)^{-3(w+\delta)}}
{1+\left(\tfrac{\Omega_{\text{(dm,0)}}}{\Omega_{\text{(de,0)}}}\right)}\right]^{-\tfrac{\delta}{w+\delta}} \\[1mm]
& + \Omega_{\text{bm,0}}(1+z)^3 
+ \Omega_{\text{r,0}}(1+z)^4 \Bigg\}^{\tfrac{1}{2}}.
\end{split}   
\end{gather}

This interaction is guaranteed to always yield positive energy densities, regardless of the sign or magnitude of the interaction parameter $\delta$.

Previous observational constraints for this model may be found in~\cite{He:2008tn,vonMarttens:2018iav,Arevalo:2011hh,Aljaf:2020eqh,vanderWesthuizen:2025iam,Yang:2022csz}.

\subsubsection*{Non-linear IDE model 2: $Q=3H\delta \left( \frac{\rho_{\text{dm}}^2  }{\rho_{\text{dm}}+\rho_{\text{de}}} \right)$}
 \label{subsec:model_non_linear_2}

The normalized Hubble function for this model is found in~\cite{vanderWesthuizen:2025rip} and is given by:
\begin{gather}
\begin{split} \label{hz_Q_dmdm}
E(z) = \Bigg\{ & \Biggl[\Omega_{\text{(dm,0)}} 
+ \Omega_{\text{(de,0)}} \left(
\frac{\Bigl[w+\delta \left(\tfrac{\Omega_{\text{(dm,0)}}}{\Omega_{\text{(de,0)}}}\right)\Bigr](1+z)^{3w} 
- \delta \left(\tfrac{\Omega_{\text{(dm,0)}}}{\Omega_{\text{(de,0)}}}\right)}{w}
\right)\Biggr] 
(1+z)^{3\left(1-\tfrac{w\delta}{w-\delta}\right)} \\[1mm]
& \times \left[
\frac{\Bigl[w+\delta \left(\tfrac{\Omega_{\text{(dm,0)}}}{\Omega_{\text{(de,0)}}}\right)\Bigr](1+z)^{3w} 
+ \left(\tfrac{\Omega_{\text{(dm,0)}}}{\Omega_{\text{(de,0)}}}\right)(w-\delta)}
{w\Bigl[1+\tfrac{\Omega_{\text{(dm,0)}}}{\Omega_{\text{(de,0)}}}\Bigr]}
\right]^{\tfrac{\delta}{w-\delta}}  + \Omega_{\text{bm,0}}(1+z)^3 
+ \Omega_{\text{r,0}}(1+z)^4 \Bigg\}^{\tfrac{1}{2}}.
\end{split}   
\end{gather}
If energy flows from dark matter to dark energy (given by $\delta<0$), this interaction leads to negative dark energy densities in the past, while all energy densities remain positive for a small interaction from dark energy to dark matter (given by $\delta>0$).

Previous observational constraints for this model may be found in~\cite{vonMarttens:2018iav,Arevalo:2011hh,Aljaf:2020eqh}.

\subsubsection*{Non-linear IDE model 3: $Q=3H\delta \left( \frac{\rho_{\text{de}}^2  }{\rho_{\text{dm}}+\rho_{\text{de}}} \right)$}
 \label{subsec:model_non_linear_3}

The normalized Hubble function for this model is found in~\cite{vanderWesthuizen:2025rip} and is given by:
\begin{gather}
\begin{split} \label{hz_Q_dede}
E(z) = \Bigg\{ & \Biggl[\Omega_{\text{(dm,0)}} 
\left(\frac{\Bigl[w \left( \tfrac{\Omega_{\text{(dm,0)}}}{\Omega_{\text{(de,0)}}}\right)+\delta \Bigr](1+z)^{-3w}-\delta}{w \left( \tfrac{\Omega_{\text{(dm,0)}}}{\Omega_{\text{(de,0)}}}\right)}\right) 
+ \Omega_{\text{(de,0)}}\Biggr] (1+z)^{3\left(1+\tfrac{w^2}{w-\delta}\right)} \\[1mm]
& \times \left[\frac{\Bigl[w \left( \tfrac{\Omega_{\text{(dm,0)}}}{\Omega_{\text{(de,0)}}}\right)+\delta \Bigr](1+z)^{-3w}+w-\delta}{w \left(1+ \tfrac{\Omega_{\text{(dm,0)}}}{\Omega_{\text{(de,0)}}}\right)}\right]^{\tfrac{\delta}{w-\delta}} + \Omega_{\text{bm,0}}(1+z)^3 
+ \Omega_{\text{r,0}}(1+z)^4 \Bigg\}^{\tfrac{1}{2}}.
\end{split}   
\end{gather}

If energy flows from dark matter to dark energy (given by $\delta<0$), this interaction predicts negative dark matter densities in the future, while all energy densities remain positive for a small interaction from dark energy to dark matter (given by $\delta>0$).

Previous observational constraints for this model may be found in~\cite{vonMarttens:2018iav,Arevalo:2011hh,Aljaf:2020eqh,Lima:2024wmy}.

\subsection{Theoretical constraints on the models} \label{theo_const}

The eight interaction kernels introduced in the previous section may lead to imaginary and/or undefined energy densities for either dark matter or dark energy. The conditions required to avoid such pathological situations are summarized in Table~\ref{tab:Com_real}. Similarly, these models may exhibit negative dark matter or dark energy densities in the past or future, as well as possible future big rip singularities, if the conditions listed in Table~\ref{tab:Com_PEC} are not satisfied.
It is worth noting that for seven of the eight interactions studied here, negative energy densities are an inevitable consequence of energy flow from dark matter to dark energy. Predictions of future negative energy densities and future singularities are less problematic than negative energy densities in the past, since we have no observational data probing the future expansion of the Universe, and such features may simply signal a breakdown of the model beyond its domain of applicability. In contrast, the presence of negative energy densities in the past represents a more serious issue, as it would violate several energy conditions of general relativity, including the Weak Energy Condition (WEC). 
For discussions on the violation of energy conditions, see~\cite{Carroll:2003st,Santos:2007pp,Rubakov:2014jja,Martin-Moruno:2017exc}, while discussions on the possibility of negative dark energy, and how this may be useful in addressing recent problems in cosmology, can be found in~\cite{Valiviita:2009nu,BOSS:2014hwf,Poulin:2018zxs,Wang:2018fng,Visinelli:2019qqu,Calderon:2020hoc,Ong:2022wrs,Malekjani:2023ple,Bouhmadi-Lopez:2025spo,Gomez-Valent:2025mfl,Gonzalez-Fuentes:2025lei,Ghafari:2025eql}.

\begin{table}[h]
\centering
\renewcommand{\arraystretch}{1.2} % Adjust row spacing
\setlength{\tabcolsep}{10pt}     % Adjust column spacing
\begin{tabular}{|c|c|c|}
\hline
\textbf{Interaction $Q$} 
 & Conditions to avoid imaginary $\rho_{\text{dm/de}}$
 & Conditions to avoid undefined $\rho_{\text{dm/de}}$ \\ \hline \hline

$3 H (\delta_{\text{dm}} \rho_{\text{dm}} + \delta_{\text{de}}  \rho_{\text{de}})$
 & $(\delta_{\text{dm}}+\delta_{\text{de}}+w)^2>4\delta_{\text{de}}\delta_{\text{dm}}$
 & $w\ne0$ ; $(\delta_{\text{dm}}+\delta_{\text{de}}+w)^2-4\delta_{\text{de}}\delta_{\text{dm}} \neq 0$
\\ \hline

$3H\delta( \rho_{\text{dm}}+\rho_{\text{de}})$
 & $\delta\le-\frac{w}{4}$
 & $w\ne0$ ; $\delta\ne-\frac{w}{4}$
\\ \hline

$3H\delta( \rho_{\text{dm}}-\rho_{\text{de}})$
 & $\rho_{\text{dm/de}}$ \text{ always real}
 & $w\ne0$
\\ \hline

$3H \delta \rho_{\text{dm}}$
 & $\rho_{\text{dm/de}}$ \text{ always real}
 & $\delta\ne -w$
\\ \hline

$3H\delta \rho_{\text{de}}$
 & $\rho_{\text{dm/de}}$ \text{ always real}
 & $\delta\ne -w$
\\ \hline

$3H\delta \left( \frac{\rho_{\text{dm}} \rho_{\text{de}} }{\rho_{\text{dm}}+\rho_{\text{de}}} \right)$
 & $\rho_{\text{dm/de}}$ \text{ always real}
 & $\delta\ne -w$
\\ \hline

$3H\delta \left( \frac{\rho^2_{\text{dm}} }{\rho_{\text{dm}}+\rho_{\text{de}}} \right)$
 & $\rho_{\text{dm/de}}$ \text{ always real}
 & $w<0$ ; $w<\delta\le-\frac{w}{r_0}$
\\ \hline

$3H\delta \left( \frac{\rho^2_{\text{de}} }{\rho_{\text{dm}}+\rho_{\text{de}}} \right)$
 & $\rho_{\text{dm/de}}$ \text{ always real}
 &  $w<0$ ; $w<\delta\le-w r_0$
\\ \hline

\end{tabular}
\caption{Conditions required to avoid imaginary or undefined energy densities for the different interaction kernels, as derived in~\cite{vanderWesthuizen:2025vcb, vanderWesthuizen:2025mnw}. Here, we define $r_0 = \left( \tfrac{\Omega_{\text{(dm,0)}}}{\Omega_{\text{(de,0)}}} \right)$.}
\label{tab:Com_real}
\end{table}

\begin{table}[h]
\centering
\renewcommand{\arraystretch}{1.1} % Adjust row spacing
\setlength{\tabcolsep}{10pt}     % Adjust column spacing
\begin{tabular}{|c|c|c|c|}
\hline
\textbf{Interaction $Q$} 
 & \text{$\rho_{\text{dm/de}}>0$ domain} 
 & \text{$\rho_{\text{dm/de}}>0$ conditions} 
 & \text{No future big rip}
\\ \hline \hline

$3 H (\delta_{\text{dm}} \rho_{\text{dm}} + \delta_{\text{de}}  \rho_{\text{de}})$
 & \text{DE $\rightarrow$ DM}
 & $\; \delta_{\text{dm}}\ge0 ; \; \delta_{\text{de}}\ge0 ; \;\delta_{\text{dm}}r_0+ \delta_{\text{de}}\le-\frac{w r_0}{(1+r_0)}$ & $ \delta_{\text{dm}} \left(w+1\right)-\delta_{\text{de}} \le w + 1$
\\ \hline

$3H\delta( \rho_{\text{dm}}+\rho_{\text{de}})$
 & \text{DE $\rightarrow$ DM}
 & $0 \le \delta \le -\frac{w r_0}{(1+r_0)^2}$
 & $\delta\ge 1+\frac{1}{ w}$
\\ \hline

$3H\delta( \rho_{\text{dm}}-\rho_{\text{de}})$
 & \text{No viable domain}
 & \text{No viable domain}
 & $\delta \le \frac{1 + w}{2+w}$
\\ \hline

$3H \delta \rho_{\text{dm}}$
 & \text{DE $\rightarrow$ DM}
 & $0 \leq \delta \leq -\frac{w}{(1 + r_0)}$
 & $w>-1$
\\ \hline

$3H\delta \rho_{\text{de}}$
 & \text{DE $\rightarrow$ DM}
 & $0 \leq \delta \leq -\frac{w}{\left(1+\frac{1}{r_0}\right)}$
 & $\delta \geq -w-1 $
\\ \hline

$3H\delta \left( \frac{\rho_{\text{dm}} \rho_{\text{de}} }{\rho_{\text{dm}}+\rho_{\text{de}}} \right)$
 & \text{DE $\leftrightarrow$ DM}
 & $\forall \delta$
 & $w\ge-1$
\\ \hline

$3H\delta \left( \frac{\rho^2_{\text{dm}} }{\rho_{\text{dm}}+\rho_{\text{de}}} \right)$
 & \text{DE $\rightarrow$ DM}
 & $0 \leq \delta \leq -\frac{w}{r_0}$
& $w\ge-1$
\\ \hline

$3H\delta \left( \frac{\rho^2_{\text{de}} }{\rho_{\text{dm}}+\rho_{\text{de}}} \right)$
 & \text{DE $\rightarrow$ DM}
 & $0 \leq \delta \leq -w r_0$
& $\delta \geq  w(w+1)  $
\\ \hline

\end{tabular}
\caption{Conditions required to ensure positive energy densities and avoid future big rip singularities for the different interaction kernels. Here, $r_0 = \left( \tfrac{\Omega_{\text{(dm,0)}}}{\Omega_{\text{(de,0)}}} \right)$. These conditions are derived in~\cite{vanderWesthuizen:2025vcb, vanderWesthuizen:2025mnw}.}
\label{tab:Com_PEC}
\end{table}

An additional well-known pathology associated with interacting dark energy models is the presence of non-adiabatic instabilities in the dark energy perturbation equations. The standard treatment of this issue is to consider the doom factor $\textbf{d}$ proposed in~\cite{Gavela:2009cy}, whose sign determines the presence of early-time instabilities:
\begin{gather} \label{DSA.doom}
\begin{split}
\textbf{d} = \frac{Q}{3H\rho_{\rm{de}}(1+w)} \quad ; \quad \text{if } &\begin{cases}
\textbf{d} > 1 \quad \rightarrow \quad \text{Unstable runaway regime} \\
\textbf{d} < 0 \quad \rightarrow \quad \text{a priori free of instabilities}
\end{cases}
\end{split}
\end{gather}
Additionally, it has been pointed out that $w=-1$ causes gravitational instabilities~\cite{Lucca:2020zjb,Lucca:2021dxo}.
From these considerations, the standard approach is to identify two stable regimes in which $Q$ and $(1+w)$ have opposite signs, i.e. energy flow from dark energy to dark matter ($Q>0$) together with $w<-1$, or energy flow from dark matter to dark energy ($Q<0$) together with $w>-1$ (see Refs.~\cite{Salvatelli:2013wra,Bachega:2019fki,DiValentino:2020vnx,Lucca:2021dxo,Yang:2021hxg,Forconi:2023hsj,Giare:2024ytc,Sabogal:2024yha}).
This situation is further complicated by the fact that, for seven of the eight models considered here, we additionally find $\rho_{\rm{de}}<0$ at early times when energy flows from dark matter to dark energy ($Q<0$)~\cite{vanderWesthuizen:2025rip}. Assuming that Eq.~\eqref{DSA.doom} remains valid in the presence of negative dark energy, we instead require energy flow from dark matter to dark energy ($Q<0$ and $\rho_{\rm{de}}<0$) to be combined with a phantom dark energy equation of state, $w<-1$, in order to obtain $\textbf{d}<0$. We summarize how the stability of our models can be inferred from the posterior distributions in Table~\ref{tab:doom}.

The arguments presented above represent only one approach used in the literature. Another approach that has gained popularity in recent years involves treating perturbations within the Parametrized Post-Friedmann framework~\cite{Li:2014eha, Li:2014cee, Skordis:2015yra, Zhang:2017ize, Feng:2018yew, Dai:2019vif, Li:2020gtk, Li:2023fdk}, which is expected to open up the full parameter space. As this remains an active area of research (see~\cite{Silva:2025bnn} for recent discussions), and since the validity of using \eqref{DSA.doom} when $\rho_{\rm{de}}<0$ requires further investigation, we consider our conclusions regarding the stability of our models to be preliminary. A future work, in which we will address these issues and provide explicit perturbation equations for each of the eight models discussed here, is currently in preparation. This will allow us to perform more in-depth constraints on these models, including CMB data.

Finally, for this study we do not impose artificial constraints on the parameter space to ensure positive energy densities or stability. Instead, we allow the parameters to explore the full space, find their preferred values, and then interpret the implications of the resulting constraints for each model. The only constraints we apply are those listed in Table~\ref{tab:priors_all_models}, which are imposed to avoid undefined energy densities, as summarized in Table~\ref{tab:Com_real}, and which could otherwise cause the MCMC runs to fail.
Our approach corresponds to the $iw$CDM regime outlined in Section~4 of~\cite{vanderWesthuizen:2025rip}, where alternative approaches for setting priors that avoid the aforementioned theoretical pathologies are also discussed. As an example, recent constraints on the same linear interactions studied here---but with the direction of energy flow and the sign of $(1+w)$ chosen to ensure stability---can be found in~\cite{Costa:2016tpb}. Additionally, recent constraints on three of the linear models considered here, in which the dark matter equation of state is allowed to vary ($w_{\rm dm}\neq0$), are presented in~\cite{Wu:2025vrl}.

\subsection{Phantom-crossing, effective equations of state  $w_{\rm{dm}}^{\rm{eff}}$, $w_{\rm{de}}^{\rm{eff}}$ and the reconstructed equation of state $\tilde{w}$}
\label{phantom_background}

A topic of recent interest is whether alternatives to the $\Lambda$CDM model may allow the dark energy equation of state $w$ to cross the phantom divide $w=-1$ at some point in the past. This behaviour was hinted at for simple Chevallier--Polarski--Linder (CPL) parametrizations of dark energy, characterized by the parameters $w_0$ and $w_a$, by the DESI collaboration. In particular, DESI results suggest $w>-1$ at low redshift and $w<-1$ at higher redshift~\cite{DESI:2025fii}. For the interacting dark energy models considered here, we assume the dark energy equation of state $w$ to be constant; however, the effects of the energy exchange may still be captured through alternative descriptions.

A commonly adopted approach is to introduce effective equations of state for both dark matter, $w_{\rm{dm}}^{\rm{eff}}(z)$, and dark energy, $w_{\rm{de}}^{\rm{eff}}(z)$. This parametrization treats the system as if no interaction were present, while the evolution of each dark component is instead fully described by a redshift-dependent equation of state. The expressions for $w_{\rm{dm}}^{\rm{eff}}(z)$ and $w_{\rm{de}}^{\rm{eff}}(z)$ are obtained directly from the conservation equations~\eqref{eq:conservation} as:
\begin{gather} \label{omega_eff_dm_de}
\begin{split}
w^{\rm{eff}}_{\rm{dm}} = - \frac{Q}{3 H \rho_{\rm{dm}}} \qquad \text{and} \qquad
w^{\rm{eff}}_{\rm{de}} = w_{\rm{de}} + \frac{Q}{3 H \rho_{\rm{de}}}.
\end{split}
\end{gather}
Intuitively, one can note that $\rho_{\rm{de}}$ decreases when $w^{\rm{eff}}_{\rm{de}}>-1$ (the quintessence regime), while $\rho_{\rm{de}}$ increases when $w^{\rm{eff}}_{\rm{de}}<-1$ (the phantom regime).

Another approach is to introduce the reconstructed dark energy equation of state $\tilde{w}(z)$, which can be obtained from the normalized Hubble parameter $E(z)$ in a flat universe in the absence of interactions in the dark sector, such that:
\begin{gather} \label{wz_1}
\begin{split}
E^2(z)&= \Omega_{\text{(r,0)}}(1+z)^{4}+ \Omega_{\text{(bm,0)}}(1+z)^{3}+ \Omega_{\text{(dm,0)}}(1+z)^{3}+  \Omega_{\text{(de,0)}} \text{exp}\left[ 3 \int_0^z dz' \frac{1+\tilde{w}(z')}{1 + z'} \right], \\
\end{split}
\end{gather}
From \eqref{wz_1}, a general expression for $\tilde{w}(z)$ can be obtained for any interacting dark energy model that satisfies the same conservation equation~\eqref{eq:conservation}. As derived in Appendix~C of~\cite{vanderWesthuizen:2025mnw}, this expression is given by:
\begin{gather} \label{wz_2}
\begin{split}
\tilde{w}(z)&=  \frac{w(z)\rho_{\text{de}}}{\rho_{\text{dm}}+\rho_{\text{de}}-\rho_{\text{(dm,0)}}(1+z)^{3}}.
\end{split}
\end{gather}
where $\rho_{\text{dm}}$ and $\rho_{\text{de}}$ are the dark matter and dark energy densities of the model under consideration. In the $\Lambda$CDM limit, one recovers $\tilde{w}(z)=w$.

In general, we find that when energy flows from dark energy to dark matter, a divergent phantom-crossing behaviour is observed for $\tilde{w}(z)$~\cite{Gavela:2009cy, vanderWesthuizen:2025mnw}. This divergent behaviour is not pathological, but rather an artefact of the parametrization.
For illustrative purposes, the evolution of $w^{\rm{eff}}_{\rm{dm}}(z)$, $w^{\rm{eff}}_{\rm{de}}(z)$, and $\tilde{w}(z)$ for all eight models, using the mean parameter values obtained from the analysis, is shown in Fig.~\ref{fig:w_eff}.

\section{Observational constraints}\label{sec:method}

Once the set of models has been described, we aim to determine the values of the coupling parameters in each case, as well as the remaining cosmological parameters in these scenarios. We use the public Markov Chain Monte Carlo code \texttt{MontePython}~\cite{Brinckmann:2018cvx,Audren:2012wb}, together with modified versions of \texttt{CLASS}~\cite{Blas:2011rf} tailored to each interacting scenario. We consider the following datasets:
\begin{itemize}[noitemsep]
\item Pantheon+ Type Ia supernova data~\cite{Brout:2022vxf}, calibrated with SH0ES Cepheid data.
\item DESI DR2 Baryon Acoustic Oscillation data~\cite{DESI:2025zgx}.
\item Cosmic Clock measurements~\cite{Moresco:2024wmr}.
\item Big Bang Nucleosynthesis as a prior on $\Omega_b h^2 = 0.02218 \pm 0.00055$~\cite{Schoneberg:2024ifp,Burns:2023sgx}\footnote{The same prior is used in the DESI DR2 analysis~\cite{DESI:2025zgx} to break the well-known degeneracy in $H_0 r_d$.}.
\end{itemize}
In our analyses, we consider as cosmological parameters the fractional density of dark matter $\Omega_{\rm dm}$, the fractional density of baryons $\Omega_{\rm b}$, the Hubble constant $H_0$, and the dark energy equation of state $w$. In each model, we treat the interaction parameters ($\delta$, $\delta_{\rm dm}$ and/or $\delta_{\rm de}$) as free parameters, while the total matter density $\Omega_{\rm m}$ is taken as a derived parameter. Due to the inclusion of Pantheon+ data, we also include the absolute magnitude of Type Ia supernovae, $M$, as a nuisance parameter.
We assume three massive neutrinos with a total mass $\sum m_\nu = 0.06\,\mathrm{eV}$ and an effective number of relativistic species $N_{\rm eff} = 3.044$, while all other parameters are set to their default values in \texttt{CLASS}. We use flat priors with bounds summarized in Table~\ref{tab:priors_all_models} for each parameter.
The interaction parameters are assigned unbounded flat priors, $\delta \in (-\infty,+\infty)$, except for the non-linear model given by $Q=3H\delta \left( \frac{\rho_{\text{dm}}^2}{\rho_{\text{dm}}+\rho_{\text{de}}} \right)$, for which we impose $\delta \in [w,-0.8w/r_0]$, and the non-linear model given by $Q=3H\delta \left( \frac{\rho_{\text{de}}^2}{\rho_{\text{dm}}+\rho_{\text{de}}} \right)$, for which we impose $\delta \in [w,-0.8w r_0]$. These bounds are required to avoid undefined dark matter or dark energy densities, in particular divisions by zero that cause $E(z)$ to diverge as the parameter space approaches the limits listed in Table~\ref{tab:Com_real}.
As also shown in Table~\ref{tab:Com_real}, additional conditions can be imposed to avoid both negative energy densities and future big rip singularities. However, we do not enforce these constraints, as our goal is to investigate whether regions of parameter space associated with such features may be preferred by the data. This approach may, for example, indicate that negative energy densities are favoured in certain models.

\begin{table}[h!]
\centering
\resizebox{\textwidth}{!}{%
\begin{tabular}{lccccccc}
\hline
Model / Kernel $Q$ &
$\boldsymbol{\Omega_{\rm dm}}$ &
$\boldsymbol{\Omega_{\rm b}}$ &
$\boldsymbol{H_0}$ &
$\boldsymbol{M}$ &
$\boldsymbol{w}$ &
$\boldsymbol{\delta_{\rm dm}}$ ~or~ $\boldsymbol{\delta}$ &
$\boldsymbol{\delta_{\rm de}}$  ~or~ $\boldsymbol{\delta}$  \\
\hline
%-------------------------------------------
$\Lambda$CDM &
$[0.001,\,0.9]$ &
$[0.001,\, 0.3]$ &
$[40,\,100]$ &
$[-30,\,-10]$ &
-- &
-- &
-- \\
%-------------------------------------------
$3H(\delta_{\rm dm}\rho_{\rm dm} + \delta_{\rm de}\rho_{\rm de})$ &
$[0.001,\,0.9]$ &
$[0.001,\, 0.3]$ &
$[40,\,100]$ &
$[-30,\,-10]$ &
$[-2.00,\,-0.33]$ &
$(-\infty,\,+\infty)$ &
$(-\infty,\,+\infty)$ \\
%-------------------------------------------
$3H\delta(\rho_{\rm dm}+\rho_{\rm de})$ &
$[0.001,\,0.9]$ &
$[0.001,\, 0.3]$ &
$[40,\,100]$ &
$[-30,\,-10]$ &
$[-2.00,\,-0.33]$ &
$(-\infty,\,+\infty)$ &
-- \\
%-------------------------------------------
$3H\delta(\rho_{\rm dm}-\rho_{\rm de})$ &
$[0.001,\,0.9]$ &
$[0.001,\, 0.3]$ &
$[40,\,100]$ &
$[-30,\,-10]$ &
$[-2.00,\,-0.33]$ &
$(-\infty,\,+\infty)$ &
-- \\
%-------------------------------------------
%$3H\delta_{\rm dm}\rho_{\rm dm}$ &
$3H\delta\rho_{\rm dm}$ &
$[0.001,\,0.9]$ &
$[0.001,\, 0.3]$ &
$[40,\,100]$ &
$[-30,\,-10]$ &
$[-2.00,\,-0.33]$ &
$(-\infty,\,+\infty)$ &
-- \\
%-------------------------------------------
%$3H\delta_{\rm de}\rho_{\rm de}$ &
$3H\delta\rho_{\rm de}$ &
$[0.001,\,0.9]$ &
$[0.001,\, 0.3]$ &
$[40,\,100]$ &
$[-30,\,-10]$ &
$[-2.00,\,-0.33]$ &
-- &
$(-\infty,\,+\infty)$ \\
%-------------------------------------------
$3H\delta\!\left(\frac{\rho_{\rm dm}\rho_{\rm de}}{\rho_{\rm dm}+\rho_{\rm de}}\right)$ &
$[0.001,\,0.9]$ &
$[0.001,\, 0.3]$ &
$[40,\,100]$ &
$[-30,\,-10]$ &
$[-2.00,\,-0.33]$ &
$(-\infty,\,+\infty)$ &
-- \\
%-------------------------------------------
%$3H\delta_{\rm dm}\left(\frac{\rho_{\rm dm}^2}{\rho_{\rm dm}+\rho_{\rm de}}\right)$ &
$3H\delta
\left(\frac{\rho_{\rm dm}^2}{\rho_{\rm dm}+\rho_{\rm de}}\right)$ &
$[0.001,\,0.9]$ &
$[0.001,\, 0.3]$ &
$[40,\,100]$ &
$[-30,\,-10]$ &
$[-2.00,\,-0.33]$ &
$\big[w,\,-0.8\,w/r_0\big]$ &
-- \\
%-------------------------------------------
%$3H\delta_{\rm de}\left(\frac{\rho_{\rm de}^2}{\rho_{\rm dm}+\rho_{\rm de}}\right)$ &
$3H\delta
\left(\frac{\rho_{\rm de}^2}{\rho_{\rm dm}+\rho_{\rm de}}\right)$ &
$[0.001,\,0.9]$ &
$[0.001,\, 0.3]$ &
$[40,\,100]$ &
$[-30,\,-10]$ &
$[-2.00,\,-0.33]$ &
-- &
$\big[w,\,-0.8\,w\,r_0\big]$ \\
\hline
\end{tabular}}
\caption{\textbf{Flat priors used for all cosmological and coupling parameters} for 
$\Lambda$CDM and the interacting dark energy models.  
The coupling parameters $\delta$, $\delta_{\rm dm}$, and $\delta_{\rm de}$ are assigned unbounded 
flat priors, except for the last two non-linear models, where bounds are imposed to avoid undefined densities, as summarized in Table~\ref{tab:Com_real}.  
Here, $r_0 = \Omega_{\rm dm,0}/(1-\Omega_{\rm m,0})$.}
\label{tab:priors_all_models} 
\end{table}

\subsection{Linear models}
 
\begin{figure}[h!]
\centering
\includegraphics[width=1\linewidth]{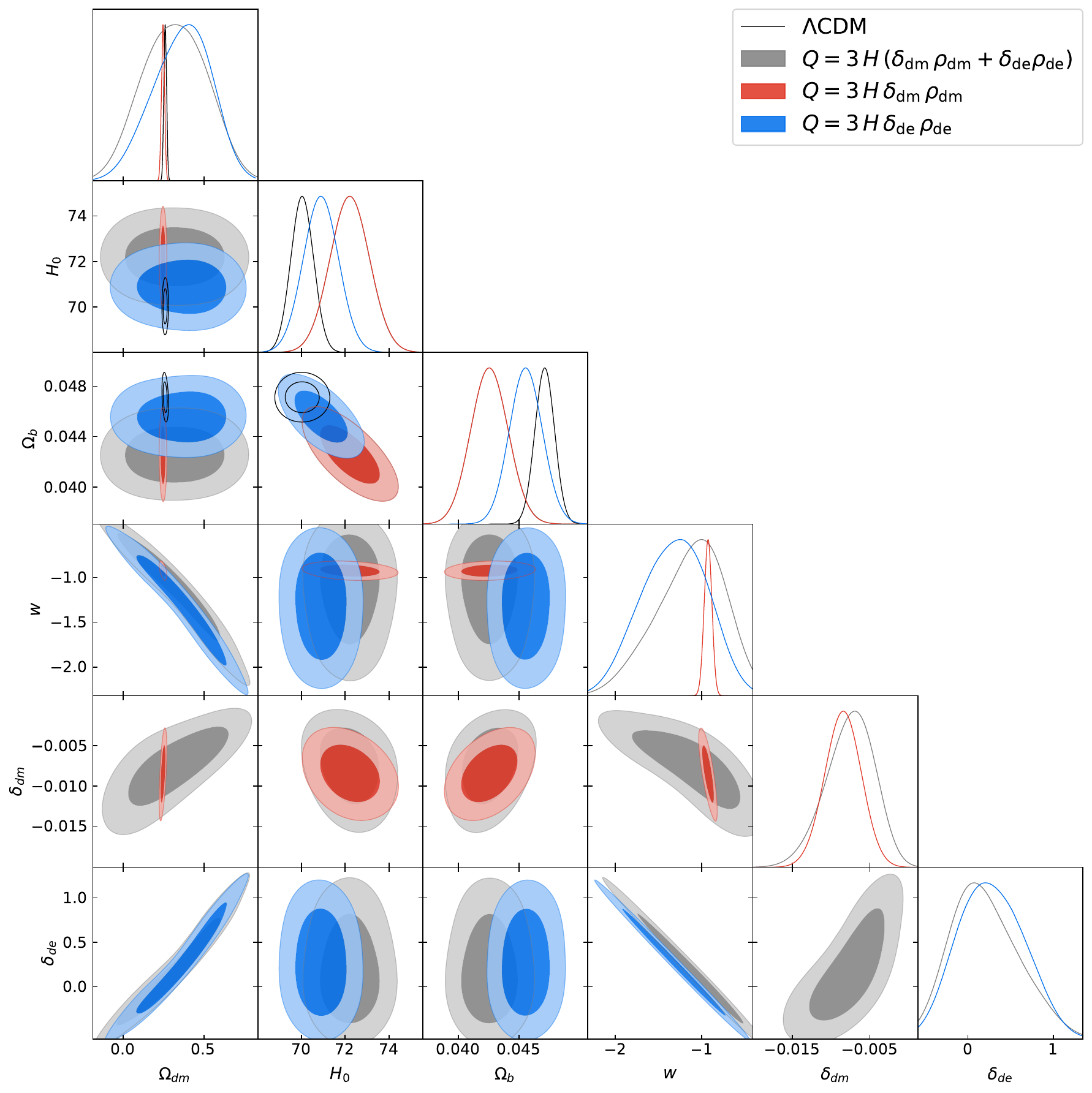}
\caption{\textbf{Pantheon+, DESI DR2, Cosmic Clocks \& BBN} – One-dimensional posterior distributions and two-dimensional contours obtained for several parameters for the reference $\Lambda$CDM model and the models given by $Q = 3\,H\,\delta\,\rho_{\mathrm{dm}}$ and $Q = 3\,H\,\delta\,\rho_{\mathrm{de}}$, discussed in Section~\ref{subsec:model_linear}.}
\label{fig:model_P-D-C-B_2-4_2-5}
\end{figure}

We begin by considering the most general linear model in which both interaction parameters, $\delta_{\rm dm}$ and $\delta_{\rm de}$, are allowed to vary ($Q = 3H(\delta_{\text{dm}}\rho_{\text{dm}} + \delta_{\text{de}}\rho_{\text{de}})$), as well as the cases in which only one of them is allowed to vary ($Q = 3H\delta\rho_{\text{dm}}$ and $Q = 3H\delta\rho_{\text{de}}$). These models were introduced in Section~\ref{subsec:model_linear}.
In Table~\ref{tab:constraints_pantheon_desi}, we present the results obtained using Pantheon+ and DESI DR2 data, while Table~\ref{tab:constraints_all_models} shows the corresponding results when additional information from Cosmic Clocks and BBN is also incorporated. For a clearer visualization, the contour plots corresponding to the latter case are shown in Fig.~\ref{fig:model_P-D-C-B_2-4_2-5}.

\underline{Takeaways from MCMC analysis for $Q=3 H (\delta_{\text{dm}} \rho_{\text{dm}} + \delta_{\text{de}}  \rho_{\text{de}})$, $Q=3H \delta \rho_{\text{dm}}$, and $Q=3H \delta \rho_{\text{de}}$:}

\begin{itemize}
\item A detection of the interaction sourced by $\delta_{\rm dm}$ is found, while the constraining power for the interaction parameter $\delta_{\rm de}$ remains very poor. This lack of sensitivity to $\delta_{\rm de}$, despite obtaining consistent results for $\delta_{\rm dm}$, persists even when both interactions are allowed to act simultaneously. This indicates that the data do not provide sufficient constraining power to isolate the effects of $\delta_{\rm de}$, since at the background level an interaction proportional to $\rho_{\rm de}$ largely mimics a dynamical dark energy expansion history and is therefore highly degenerate with non-interacting DDE models when using background observables alone, whereas an interaction proportional to $\rho_{\rm dm}$ directly modifies the matter dilution rate and can be more effectively constrained.

\item Since $\delta_{\rm dm}<0$, the models predict energy flow from dark matter to dark energy and the presence of negative dark energy at some point in the past. Since $\delta_{\rm de}$ is not well constrained, negative values are still allowed, which could in turn predict negative dark matter densities in the future.

\item The matter content is strongly affected by the nature of the interaction. When both interactions are simultaneously active, or when only $\delta_{\rm de}$ is allowed to vary, a large increase in the total matter density is observed. Conversely, when only $\delta_{\rm dm}$ is allowed to vary, a mild decrease in the matter density is found. 

\item In addition, phantom-like dark energy behaviour appears whenever the interaction parameter $\delta_{\rm de}$ is included, either alone or in combination with $\delta_{\rm dm}$. This points toward an effective equation of state $w<-1$, although such an effect must be interpreted cautiously, since it may simply arise from parameter degeneracies involving $w$ and $\delta_{\rm de}$. For the interaction $Q=3H\delta \rho_{\rm{de}}$, dark energy has an effective equation of state $w^{\rm{eff}}_{\rm{de}} = w+\delta$, as obtained from \eqref{omega_eff_dm_de}. This implies that, even if $w<-1$, one may still have $w^{\rm{eff}}_{\rm{de}}>-1$, thereby avoiding consequences such as a future big rip singularity. This behaviour is illustrated by the regions of parameter space inside and outside the green mesh in panel~5 of Fig.~\ref{fig:parameter_space_BBN}. See~\cite{vanderWesthuizen:2025vcb} for a detailed analysis of the consequences of $w<-1$ and big rip singularities for the linear IDE models considered here. 

\end{itemize}

\begin{figure}[h!]
\centering
\includegraphics[width=1\linewidth]{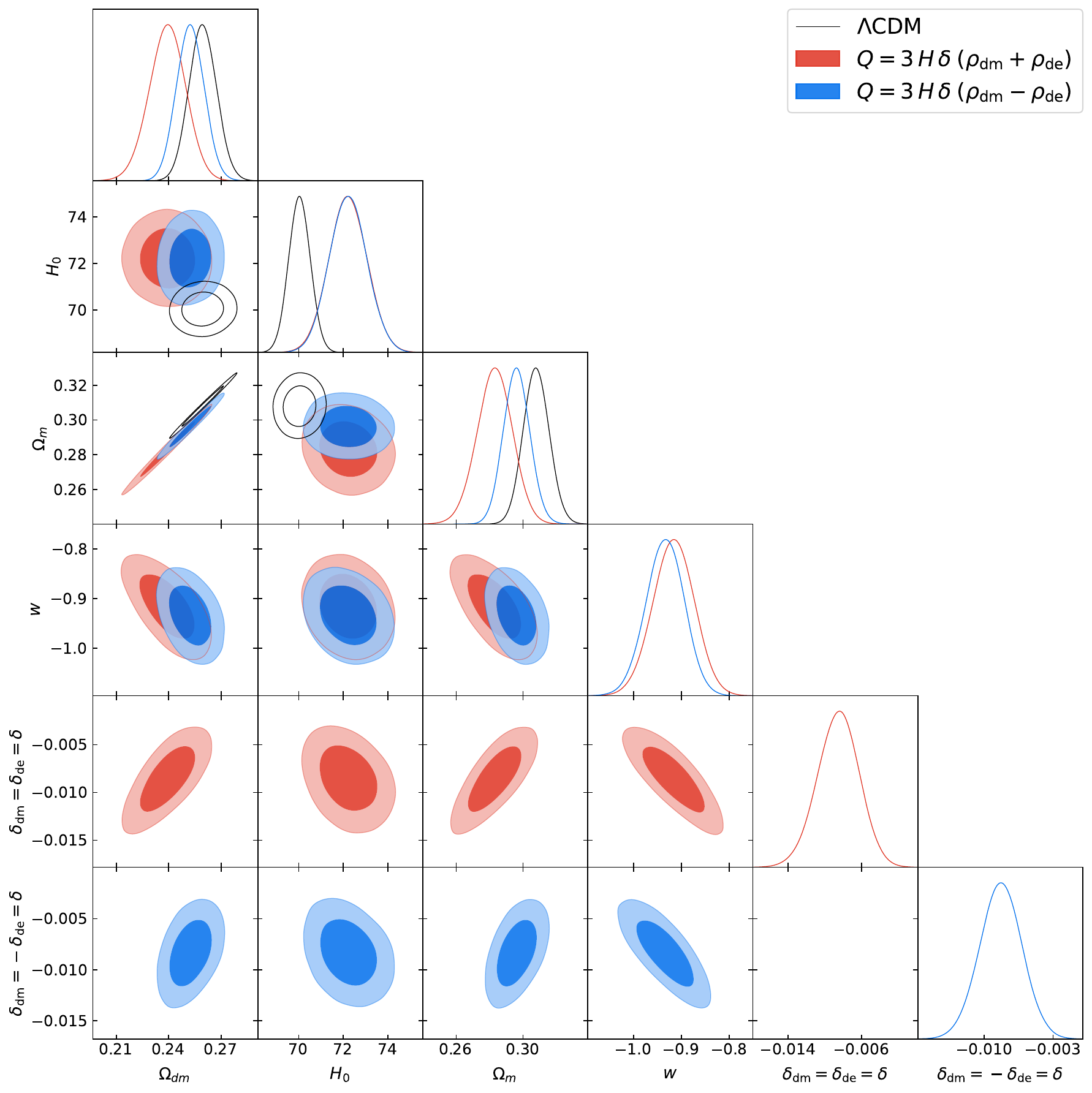}
\caption{\textbf{Pantheon+, DESI DR2, Cosmic Clocks \& BBN} – One-dimensional posterior distributions and two-dimensional contours obtained for several parameters for the reference $\Lambda$CDM model and the models given by $Q = 3\,H\,\delta\,(\rho_{\mathrm{dm}}+\rho_{\mathrm{de}})$ and $Q = 3\,H\,\delta\,(\rho_{\mathrm{dm}}-\rho_{\mathrm{de}})$, discussed in Section~\ref{subsec:model_linear}.}
\label{fig:model_P-D-C-B_2-2_2-3}
\end{figure}

We now consider the remaining linear models in which the interaction parameters satisfy the conditions $\delta_{\rm dm} = \delta_{\rm de} \equiv \delta$ ($Q = 3H\delta(\rho_{\text{dm}}+\rho_{\text{de}})$) or $\delta_{\rm dm} = -\delta_{\rm de} \equiv \delta$ ($Q = 3H\delta(\rho_{\text{dm}}-\rho_{\text{de}})$), introduced in Section~\ref{subsec:model_linear}. We display in Table~\ref{tab:constraints_pantheon_desi} the results obtained using Pantheon+ and DESI DR2 data, and in Table~\ref{tab:constraints_all_models} those obtained when additional information from Cosmic Clocks and BBN datasets is incorporated. For a clearer visualization of the behaviour of these models, we also show the contour plots for the latter case in Fig.~\ref{fig:model_P-D-C-B_2-2_2-3}.

\underline{Takeaways from MCMC analysis for $Q=3H\delta( \rho_{\text{dm}}+\rho_{\text{de}})$ and  $Q=3H\delta( \rho_{\text{dm}}-\rho_{\text{de}})$:}
\begin{itemize}

\item A detection of the interaction parameter $\delta$ is obtained in both cases, pointing toward a non-negligible exchange of energy between the dark sectors. Since $\delta<0$ in both cases, there is a clear preference for an initial energy flow from dark matter to dark energy, with the interaction $Q=3H\delta(\rho_{\text{dm}}-\rho_{\text{de}})$ later switching to an energy flow from dark energy to dark matter. As discussed in~\cite{vanderWesthuizen:2025vcb} and shown in Table~\ref{tab:Com_PEC}, this implies the presence of negative dark energy in the past for both models, and additionally predicts negative dark matter in the future for the interaction $Q=3H\delta(\rho_{\text{dm}}+\rho_{\text{de}})$.  
\item Regarding the matter content, we observe a mild decrease in the total matter density compared to $\Lambda$CDM, although this effect is slightly smaller for the interaction $Q=3H\delta(\rho_{\text{dm}}-\rho_{\text{de}})$.

\end{itemize}

\begin{table}[h!]
\centering
\resizebox{\textwidth}{!}{%
\begin{tabular}{lcccccc}
\hline
Model / Kernel $Q$ &
$\boldsymbol{\Omega_{\rm dm}}$ &
$\boldsymbol{H_0}$ &
$\boldsymbol{M}$ &
$\boldsymbol{w}$ &
$\boldsymbol{\delta_{\rm dm}}$ ~or~ $\boldsymbol{\delta}$ &  $\boldsymbol{\delta_{\rm de}}$  ~or~ $\boldsymbol{\delta}$ \\
\hline
%-------------------------------------------
$\Lambda$\text{CDM} &
$0.2648^{+0.0080(0.016)}_{-0.0080(0.015)}$ &
$71.15^{+0.70(1.4)}_{-0.70(1.4)}$ &
$-19.323^{+0.022(0.042)}_{-0.022(0.042)}$ &
-- & --  & -- \\
%-------------------------------------------
$3H(\delta_{\rm dm}\rho_{\rm dm} + \delta_{\rm de}\rho_{\rm de})$ &
$0.32^{+0.18(0.30)}_{-0.18(0.30)}$ &
$73.64^{+0.99(2.0)}_{-0.99(1.9)}$ &
$-19.241^{+0.028(0.056)}_{-0.028(0.055)}$ &
$-1.10^{+0.45(0.70)}_{-0.32(0.79)}$ &
$-0.0068^{+0.0032(0.0054)}_{-0.0026(0.0060)}$ & $0.20^{+0.32(0.79)}_{-0.46(0.71)}$ \\
%-------------------------------------------
$3H\delta(\rho_{\rm dm}+\rho_{\rm de})$ &
$0.238^{+0.011(0.021)}_{-0.011(0.022)}$ &
$73.63^{+0.98(1.9)}_{-0.98(1.9)}$ &
$-19.242^{+0.028(0.055)}_{-0.028(0.055)}$ &
$-0.896^{+0.045(0.087)}_{-0.045(0.089)}$ &
$-0.0077^{+0.0023(0.0042)}_{-0.0021(0.0046)}$ &
-- \\
%-------------------------------------------
$3H\delta(\rho_{\rm dm}-\rho_{\rm de})$ &
$0.2501^{+0.0083(0.017)}_{-0.0083(0.016)}$ &
$73.63^{+0.99(2.0)}_{-0.99(1.9)}$ &
$-19.242^{+0.028(0.055)}_{-0.028(0.055)}$ &
$-0.911^{+0.041(0.079)}_{-0.041(0.081)}$ &
$-0.0076^{+0.0022(0.0041)}_{-0.0020(0.0044)}$ &
-- \\
%-------------------------------------------
%$3H\delta_{\rm dm}\rho_{\rm dm}$ &
$3H\delta\rho_{\rm dm}$ &
$0.2441^{+0.0095(0.019)}_{-0.0095(0.019)}$ &
$73.63^{+0.97(1.9)}_{-0.97(1.9)}$ &
$-19.242^{+0.028(0.055)}_{-0.028(0.055)}$ &
$-0.903^{+0.043(0.083)}_{-0.043(0.085)}$ &
$-0.0077^{+0.0022(0.0042)}_{-0.0022(0.0045)}$ &
-- \\
%-------------------------------------------
%$3H\delta_{\rm de}\rho_{\rm de}$ &
$3H\delta\rho_{\rm de}$ &
$0.351^{+0.20(0.33)}_{-0.17(0.35)}$ &
$72.90^{+0.97(1.9)}_{-0.97(1.9)}$ &
$-19.282^{+0.026(0.052)}_{-0.026(0.052)}$ &
$-1.30^{+0.42(0.71)}_{-0.37(0.70)}$ &
-- &
$0.25^{+0.37(0.75)}_{-0.42(0.72)}$ \\
%-------------------------------------------
$3H\delta\!\left(\frac{\rho_{\rm dm}\rho_{\rm de}}{\rho_{\rm dm}+\rho_{\rm de}}\right)$ &
$0.074^{+0.041(0.11)}_{-0.057(0.073)}$ &
$72.93^{+0.96(1.9)}_{-0.96(1.9)}$ &
$-19.262^{+0.027(0.053)}_{-0.027(0.053)}$ &
$-0.711^{+0.10(0.18)}_{-0.077(0.20)}$ &
$-0.85^{+0.38(0.73)}_{-0.34(0.76)}$ &
-- \\
%-------------------------------------------
%$3H\delta_{\rm dm}\!\left(\frac{\rho_{\rm dm}^2}{\rho_{\rm dm}+\rho_{\rm de}}\right)$ &
$3H\delta\!\left(\frac{\rho_{\rm dm}^2}{\rho_{\rm dm}+\rho_{\rm de}}\right)$ &
$0.2467^{+0.0091(0.018)}_{-0.0091(0.018)}$ &
$73.63^{+0.98(2.0)}_{-0.98(1.9)}$ &
$-19.242^{+0.028(0.055)}_{-0.028(0.055)}$ &
$-0.908^{+0.042(0.082)}_{-0.042(0.083)}$ &
$-0.0076^{+0.0023(0.0041)}_{-0.0020(0.0044)}$ &
-- \\
%-------------------------------------------
%$3H\delta_{\rm de}\!\left(\frac{\rho_{\rm de}^2}{\rho_{\rm dm}+\rho_{\rm de}}\right)$ &
$3H\delta\!\left(\frac{\rho_{\rm de}^2}{\rho_{\rm dm}+\rho_{\rm de}}\right)$ &
$0.428^{+0.059(0.070)}_{-0.017(0.11)}$ &
$72.94^{+0.96(1.9)}_{-0.96(1.9)}$ &
$-19.270^{+0.027(0.052)}_{-0.027(0.052)}$ &
$-1.277^{+0.045(0.17)}_{-0.10(0.13)}$ &
-- &
$0.76^{+0.38(0.44)}_{-0.15(0.59)}$ \\
\hline
\end{tabular}
}
\caption{\textbf{Pantheon+ \& DESI DR2} – mean values and 68\% constraints (with 95\% limits in parentheses)
for $\Lambda$CDM and interacting dark energy models.  
\textbf{Note:} $\Omega_{\rm m} = \Omega_{\rm dm} + \Omega_b$, with $\Omega_b = 0.048$ fixed for all models.}
\label{tab:constraints_pantheon_desi}
\end{table}

\begin{table}[h!]
\centering
\resizebox{\textwidth}{!}{%
\begin{tabular}{lccccccc}
\hline
Model / Kernel $Q$ &
$\boldsymbol{\Omega_{\rm dm}}$ &
$\boldsymbol{\Omega_{\rm b}}$ &
$\boldsymbol{H_0}$ &
$\boldsymbol{M}$ &
$\boldsymbol{w}$ &
$\boldsymbol{\delta_{\rm dm}}$ ~or~ $\boldsymbol{\delta}$ &
$\boldsymbol{\delta_{\rm de}}$  ~or~ $\boldsymbol{\delta}$ \\
\hline
%-------------------------------------------
$\Lambda$\text{CDM} &
$0.2596^{+0.0079(0.016)}_{-0.0079(0.015)}$ &
$0.04713^{+0.00076(0.0015)}_{-0.00076(0.0015)}$ &
$70.04^{+0.49(0.96)}_{-0.49(0.95)}$ &
$-19.358^{+0.015(0.030)}_{-0.015(0.030)}$ &
-- & -- & -- \\
%-------------------------------------------
$3H(\delta_{\rm dm}\rho_{\rm dm} + \delta_{\rm de}\rho_{\rm de})$ &
$0.32^{+0.18(0.29)}_{-0.18(0.30)}$ &
$0.0426^{+0.0015(0.0029)}_{-0.0015(0.0028)}$ &
$72.21^{+0.85(1.7)}_{-0.85(1.7)}$ &
$-19.287^{+0.024(0.047)}_{-0.024(0.047)}$ &
$-1.13^{+0.46(0.72)}_{-0.33(0.80)}$ &
$-0.0075^{+0.0034(0.0058)}_{-0.0028(0.0064)}$ &
$0.21^{+0.33(0.80)}_{-0.47(0.72)}$ \\
%-------------------------------------------
$3H\delta(\rho_{\rm dm}+\rho_{\rm de})$ &
$0.239^{+0.010(0.020)}_{-0.010(0.021)}$ &
$0.0426^{+0.0015(0.0029)}_{-0.0015(0.0028)}$ &
$72.21^{+0.85(1.7)}_{-0.85(1.7)}$ &
$-19.287^{+0.024(0.047)}_{-0.024(0.047)}$ &
$-0.916^{+0.043(0.084)}_{-0.043(0.086)}$ &
$-0.0086^{+0.0023(0.0044)}_{-0.0023(0.0046)}$ &
-- \\
\
%-------------------------------------------
$3H\delta(\rho_{\rm dm}-\rho_{\rm de})$ &
$0.2525^{+0.0080(0.016)}_{-0.0080(0.015)}$ &
$0.0426^{+0.0014(0.0029)}_{-0.0014(0.0028)}$ &
$72.22^{+0.84(1.7)}_{-0.84(1.6)}$ &
$-19.287^{+0.024(0.046)}_{-0.024(0.046)}$ &
$-0.934^{+0.040(0.077)}_{-0.040(0.080)}$ &
$-0.0083^{+0.0022(0.0042)}_{-0.0022(0.0042)}$ &
-- \\

%-------------------------------------------
%$3H\delta_{\rm dm}\rho_{\rm dm}$ &
$3H\delta\rho_{\rm dm}$ &
$0.2459^{+0.0091(0.018)}_{-0.0091(0.018)}$ &
$0.0426^{+0.0015(0.0029)}_{-0.0015(0.0028)}$ &
$72.21^{+0.85(1.7)}_{-0.85(1.6)}$ &
$-19.287^{+0.024(0.047)}_{-0.024(0.047)}$ &
$-0.925^{+0.041(0.080)}_{-0.041(0.082)}$ &
$-0.0085^{+0.0022(0.0043)}_{-0.0022(0.0045)}$ &
-- \\
%-------------------------------------------
%$3H\delta_{\rm de}\rho_{\rm de}$ &
$3H\delta\rho_{\rm de}$ &
$0.36^{+0.20(0.33)}_{-0.16(0.35)}$ &
$0.0456^{+0.0013(0.0026)}_{-0.0013(0.0025)}$ &
$70.89^{+0.76(1.5)}_{-0.76(1.5)}$ &
$-19.339^{+0.020(0.038)}_{-0.020(0.039)}$ &
$-1.32^{+0.41(0.71)}_{-0.37(0.68)}$ &
-- &
$0.27^{+0.37(0.75)}_{-0.42(0.72)}$ \\
%-------------------------------------------
$3H\delta\!\left(\frac{\rho_{\rm dm}\rho_{\rm de}}{\rho_{\rm dm}+\rho_{\rm de}}\right)$ &
$0.083^{+0.044(0.12)}_{-0.061(0.082)}$ &
$0.0449^{+0.0012(0.0025)}_{-0.0012(0.0024)}$ &
$71.06^{+0.74(1.5)}_{-0.74(1.4)}$ &
$-19.319^{+0.021(0.041)}_{-0.021(0.041)}$ &
$-0.731^{+0.11(0.19)}_{-0.080(0.21)}$ &
$-0.78^{+0.33(0.63)}_{-0.33(0.65)}$ &
-- \\

%-------------------------------------------
%$3H\delta_{\rm dm}\left(\frac{\rho_{\rm dm}^2}{\rho_{\rm dm}+\rho_{\rm de}}\right)$ &
$3H\delta\left(\frac{\rho_{\rm dm}^2}{\rho_{\rm dm}+\rho_{\rm de}}\right)$ &
$0.2488^{+0.0087(0.017)}_{-0.0087(0.017)}$ &
$0.0426^{+0.0015(0.0029)}_{-0.0015(0.0028)}$ &
$72.22^{+0.85(1.7)}_{-0.85(1.6)}$ &
$-19.287^{+0.024(0.047)}_{-0.024(0.047)}$ &
$-0.930^{+0.041(0.079)}_{-0.041(0.081)}$ &
$-0.0084^{+0.0022(0.0043)}_{-0.0022(0.0044)}$  &
--\\
%-------------------------------------------
%$3H\delta_{\rm de}\left(\frac{\rho_{\rm de}^2}{\rho_{\rm dm}+\rho_{\rm de}}\right)$ &
$3H\delta\left(\frac{\rho_{\rm de}^2}{\rho_{\rm dm}+\rho_{\rm de}}\right)$ &
$0.426^{+0.062(0.073)}_{-0.018(0.11)}$ &
$0.0451^{+0.0013(0.0025)}_{-0.0013(0.0024)}$ &
$71.02^{+0.74(1.5)}_{-0.74(1.5)}$ &
$-19.327^{+0.020(0.039)}_{-0.020(0.039)}$ &
$-1.273^{+0.060(0.19)}_{-0.11(0.16)}$ &
-- &
$0.74^{+0.38(0.46)}_{-0.17(0.49)}$ \\
\hline
\end{tabular}}
\caption{\textbf{Pantheon+, DESI DR2, Cosmic Clocks \& BBN} – mean values and 68\% constraints
(with 95\% limits in parentheses) for the cosmological and nuisance parameters
in $\Lambda$CDM and the eight interacting dark energy models.}
\label{tab:constraints_all_models}
\end{table}

\subsection{Non-linear models}

\begin{figure}[h!]
\centering
\includegraphics[width=1\linewidth]{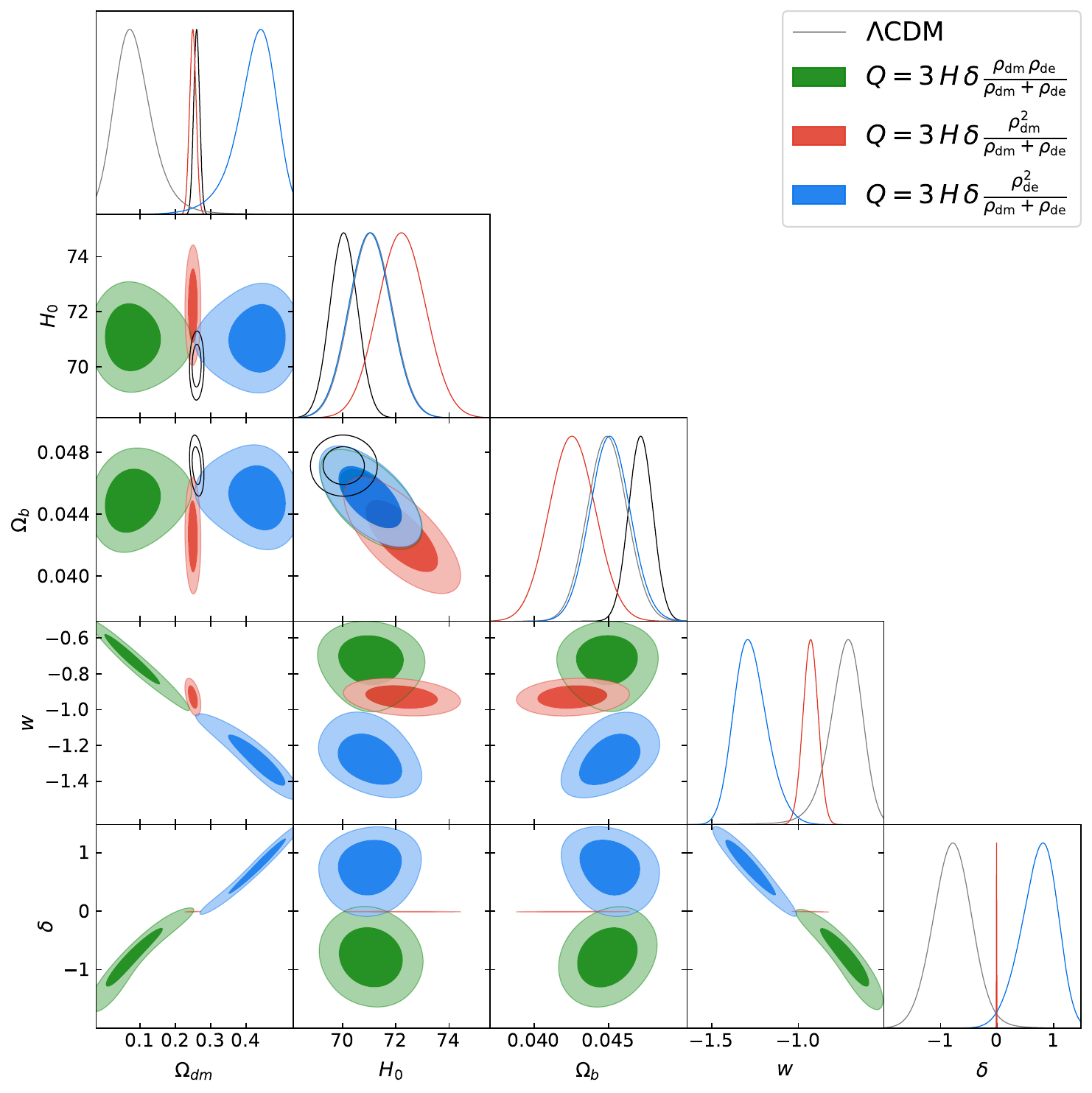}
\caption{\textbf{Pantheon+, DESI DR2, Cosmic Clocks \& BBN} – One-dimensional posterior distributions and two-dimensional contours obtained for several parameters for the reference $\Lambda$CDM model and the models given by $Q=3H\delta \left( \frac{\rho_{\text{dm}} \rho_{\text{de}} }{\rho_{\text{dm}}+\rho_{\text{de}}} \right)$, $Q=3H\delta \left( \frac{\rho^2_{\text{dm}} }{\rho_{\text{dm}}+\rho_{\text{de}}} \right)$, and $Q=3H\delta \left( \frac{\rho^2_{\text{de}} }{\rho_{\text{dm}}+\rho_{\text{de}}} \right)$, discussed in Section~\ref{subsec:model_non_linear}. }
\label{fig:model_P-D-C-B_3-2_3-3_3-4}
\end{figure}

Considering now the non-linear models presented in Section~\ref{subsec:model_non_linear}, we display in Table~\ref{tab:constraints_pantheon_desi} the results obtained using Pantheon+ and DESI DR2 data, and in Table~\ref{tab:constraints_all_models} those obtained when including the additional information from Pantheon+, DESI DR2, Cosmic Clocks, and BBN data. For a clearer visualization of these results, we also show the corresponding contour plots for the latter case in Figure~\ref{fig:model_P-D-C-B_3-2_3-3_3-4}.

\underline{Takeaways from MCMC analysis for $Q=3H\delta \left( \frac{\rho_{\text{dm}} \rho_{\text{de}} }{\rho_{\text{dm}}+\rho_{\text{de}}} \right)$, $Q=3H\delta \left( \frac{\rho^2_{\text{dm}} }{\rho_{\text{dm}}+\rho_{\text{de}}} \right)$ and $Q=3H\delta \left( \frac{\rho^2_{\text{de}} }{\rho_{\text{dm}}+\rho_{\text{de}}} \right)$:}

\begin{itemize}
  \item We find a mild detection of the three non-linear interactions from both combinations of datasets. For $Q=3H\delta \left( \frac{\rho_{\text{dm}} \rho_{\text{de}} }{\rho_{\text{dm}}+\rho_{\text{de}}} \right)$ and $Q=3H\delta \left( \frac{\rho^2_{\text{dm}} }{\rho_{\text{dm}}+\rho_{\text{de}}} \right)$, there is a preference for energy flow from dark matter to dark energy ($\delta<0$), while for $Q=3H\delta \left( \frac{\rho^2_{\text{de}} }{\rho_{\text{dm}}+\rho_{\text{de}}} \right)$ there is a preference for energy flow from dark energy to dark matter ($\delta>0$). In the case of $Q=3H\delta \left( \frac{\rho^2_{\text{dm}} }{\rho_{\text{dm}}+\rho_{\text{de}}} \right)$, since $\delta<0$, we obtain a negative dark energy density in the past, while for the other two cases the dark sector densities remain positive at all times. For $Q=3H\delta \left( \frac{\rho^2_{\text{de}} }{\rho_{\text{dm}}+\rho_{\text{de}}} \right)$, the interaction parameter $\delta$ appears to approach the upper limit for which the energy densities remain defined and positive, $\delta=-w r_0$, as seen in the last panel of Figure~\ref{fig:parameter_space_BBN}. The parameters cannot cross this boundary due to the artificial prior limits imposed to avoid these unphysical scenarios, as summarized in Table~\ref{tab:priors_all_models}.
  \item In $Q=3H\delta \left( \frac{\rho_{\text{dm}} \rho_{\text{de}} }{\rho_{\text{dm}}+\rho_{\text{de}}} \right)$, there is a large decrease in the total matter content, primarily driven by a significant reduction in the interacting dark matter component. Nevertheless, the constraints on the dark matter and total matter density parameters remain weak, since a strong degeneracy persists between $\Omega_{\rm dm}$ and the interaction parameter $\delta$, as seen in Fig.~\ref{fig:model_P-D-C-B_3-2_3-3_3-4}. Conversely, in $Q=3H\delta \left( \frac{\rho^2_{\text{de}} }{\rho_{\text{dm}}+\rho_{\text{de}}} \right)$, we observe the opposite trend: a significant increase in the matter density, carried almost entirely by a substantial rise in the interacting dark matter component. Once again, the robustness of these results is limited by the strong degeneracy between $\Omega_{\rm dm}$ and $\delta$.  
  \item In addition, phantom-like dark energy behaviour ($w<-1$) appears in the case of $Q=3H\delta \left( \frac{\rho^2_{\text{de}} }{\rho_{\text{dm}}+\rho_{\text{de}}} \right)$, similar to what is found for the linear models when $\delta_{\rm de}$ is included. This possibility must be interpreted with care, since it may be a consequence of the strong degeneracy between the dark energy equation of state $w$ and the interaction parameter $\delta$ (see Fig.~\ref{fig:model_P-D-C-B_3-2_3-3_3-4}). Even when $w<-1$, the effective dark energy equation of state $w^{\rm eff}_{\rm de}$ given by \eqref{omega_eff_dm_de} may not lie in the phantom regime. As a result, the best-fit value can be located in a region where $w^{\rm eff}_{\rm de}>-1$, and associated phantom features such as a future big rip singularity may be avoided, as illustrated in the last panel of Figure~\ref{fig:parameter_space_BBN}. 
\end{itemize}

\subsection{Statistical analysis: $\Delta\chi^{2}$ \& $\Delta$AIC } \label{sec:stats}

\begin{table}[h!]
\centering
\resizebox{0.8\textwidth}{!}{
\begin{tabular}{l|cc|cc}
\hline
 & \multicolumn{2}{c|}{\textbf{Pantheon+ \& DESI DR2}}
 & \multicolumn{2}{c}{\textbf{Pantheon+, DESI DR2, CC \& BBN}} \\
Model / Kernel $Q$
& $\Delta\chi^{2}$ & $\Delta$AIC 
& $\Delta\chi^{2}$ & $\Delta$AIC \\
\hline
$\Lambda$CDM 
& -- & -- 
& -- & -- \\[2pt]

$3H(\delta_{\rm dm}\rho_{\rm dm}+\delta_{\rm de}\rho_{\rm de})$ 
& $-21.96$ & $-15.96$
& $-17.76$ & $-11.76$ \\[2pt]

$3H\delta(\rho_{\rm dm}+\rho_{\rm de})$ 
& $-21.94$ & $-17.94$
& $-17.76$ & $-13.76$ \\[2pt]

$3H\delta(\rho_{\rm dm}-\rho_{\rm de})$ 
& $-21.96$ & $-17.96$
& $-17.74$ & $-13.74$ \\[2pt]

$3H\delta\rho_{\rm dm}$
& $-21.94$ & $-17.94$
& $-17.74$ & $-13.74$ \\[2pt]

$ 3H\delta\rho_{\rm de}$
& $-7.52$ & $-3.52$
& $-2.34$ & $+1.66$ \\[2pt]

$3H\delta\!\left(\frac{\rho_{\rm dm}\rho_{\rm de}}{\rho_{\rm dm}+\rho_{\rm de}}\right)$ 
& $-16.14$ & $-12.14$
& $-10.28$ & $-6.28$ \\[2pt]

$3H\delta
\left(\frac{\rho_{\rm dm}^2}{\rho_{\rm dm}+\rho_{\rm de}}\right)$ & $-21.94$ & $-17.94$
& $-17.76$ & $-13.76$ \\[2pt]

$3H\delta
\left(\frac{\rho_{\rm de}^2}{\rho_{\rm dm}+\rho_{\rm de}}\right)$ & $-15.02$ & $-11.02$
& $-9.46$ & $-5.46$ \\
\hline
\end{tabular}
}
\caption{
Differences in $\chi^{2}$ and in the Akaike Information Criterion (AIC) relative to $\Lambda$CDM for all interacting dark energy models considered. The interacting models include $\Delta k = 2$ additional parameters ($w$, $\delta$), except for $Q=3H(\delta_{\rm dm}\rho_{\rm dm}+\delta_{\rm de}\rho_{\rm de})$, which has $\Delta k = 3$. Negative $\Delta$AIC values indicate improved performance relative to $\Lambda$CDM.
}
\label{tab:model_statistics}
\end{table}

In order to assess the large number of models presented in this work, we compare them statistically to the reference scenario, the $\Lambda$CDM model. Our first comparison relies on the relative $\chi^2$ value with respect to $\Lambda$CDM, defined as $\Delta\chi^2 = \chi^2_{\text{model}} - \chi^2_{\Lambda\text{CDM}}$, such that negative values correspond to a better fit to the data. The relative values are displayed in Table~\ref{tab:model_statistics}.

Regarding the linear models, we find that they provide a substantial improvement in the fit, with the only exception being $Q=3H\delta\rho_{\text{de}}$, for which the preference is smaller, although still present. These results hold for both dataset combinations considered, namely Pantheon+ and DESI DR2, and Pantheon+, DESI DR2, Cosmic Clocks, and BBN. However, when Cosmic Clocks and BBN data are added, the statistical preferences are reduced, although they remain clear for all linear models, with the previously mentioned exception of $Q=3H\delta\rho_{\text{de}}$. In this specific case, the inclusion of these datasets causes the statistical preference to almost vanish.

For the non-linear models, we also observe a remarkable improvement in the fitting performance according to the $\chi^2$ criterion, particularly for $Q=3H\delta \left( \frac{\rho_{\text{dm}}^2 }{\rho_{\text{dm}}+\rho_{\text{de}}} \right)$, which behaves similarly to $Q=3H \delta \rho_{\text{dm}}$. Conversely, $Q=3H\delta \left( \frac{\rho_{\text{de}}^2 }{\rho_{\text{dm}}+\rho_{\text{de}}} \right)$, which behaves similarly to $Q=3H \delta \rho_{\text{de}}$, performs worst among the non-linear models. Once again, the inclusion of Cosmic Clocks and BBN data reduces the level of statistical preference.

However, we must take into account that we are comparing models with different numbers of parameters. In the interacting scenarios, the dark energy equation-of-state parameter is allowed to vary freely, and in addition we introduce the interaction parameter $\delta$ in each model, or the pair of interaction parameters $\delta_{\rm dm}$ and $\delta_{\rm de}$ in the case of $Q=3 H (\delta_{\text{dm}} \rho_{\text{dm}} + \delta_{\text{de}} \rho_{\text{de}})$. Consequently, these models involve two or even three additional parameters with respect to $\Lambda$CDM. It is well known that increasing the number of parameters generally leads to a better fit, but this does not necessarily imply a genuine improvement. Therefore, it is necessary to assess the statistical preference for these models by accounting for the penalty associated with the additional degrees of freedom. To this end, we rely on the Akaike Information Criterion, defined as $\mathrm{AIC} = \chi^{2} + 2\,\text{d.o.f.}$~\cite{AIC}. We define the relative difference, as before, as $\Delta\mathrm{AIC} = \mathrm{AIC}_{\text{model}} - \mathrm{AIC}_{\Lambda\text{CDM}}$ to assess the goodness of fit.

Regarding the values of $\Delta\text{AIC}$ displayed in Table~\ref{tab:model_statistics}, we find that the linear models, with the exception of $Q=3H\delta\rho_{\text{de}}$, are strongly preferred compared to the reference $\Lambda$CDM model, regardless of the dataset considered. This result, combined with the non-zero values of the coupling parameters in each model, suggests that further investigation of these scenarios is warranted. However, it should be noted that the values obtained for the coupling parameters are extraordinarily small and, as a consequence, the resulting deviations from the reference scenario are very limited.
In the specific case of $Q=3H\delta\rho_{\text{de}}$, the statistical preference is very modest when using Pantheon+ and DESI DR2 data, and it essentially vanishes once Cosmic Clocks and BBN data are included.

For the non-linear models, we also find a strong preference in all three cases when using Pantheon+ and DESI DR2 data, while this preference becomes more modest once Cosmic Clocks and BBN data are added. In this respect, the situation differs from that of the linear models. The values of the coupling parameters obtained from the constraints in each case are relatively large, implying that the associated effects may be significant and allowing these models to be distinguished from $\Lambda$CDM. In addition, we note that for $Q=3H\delta \left( \frac{\rho^2_{\text{de}} }{\rho_{\text{dm}}+\rho_{\text{de}}} \right)$ the coupling parameter $\delta$ lies close to the upper bound imposed by our priors in Table~\ref{tab:priors_all_models} to avoid undefined densities, which renders this constraint somewhat artificial compared to those obtained for the other models.

\subsection{Cosmological implications and comparison between models}
\begin{figure}[p]
    \centering

    \begin{minipage}{0.95\linewidth}
        \centering
        \includegraphics[width=0.9\linewidth]{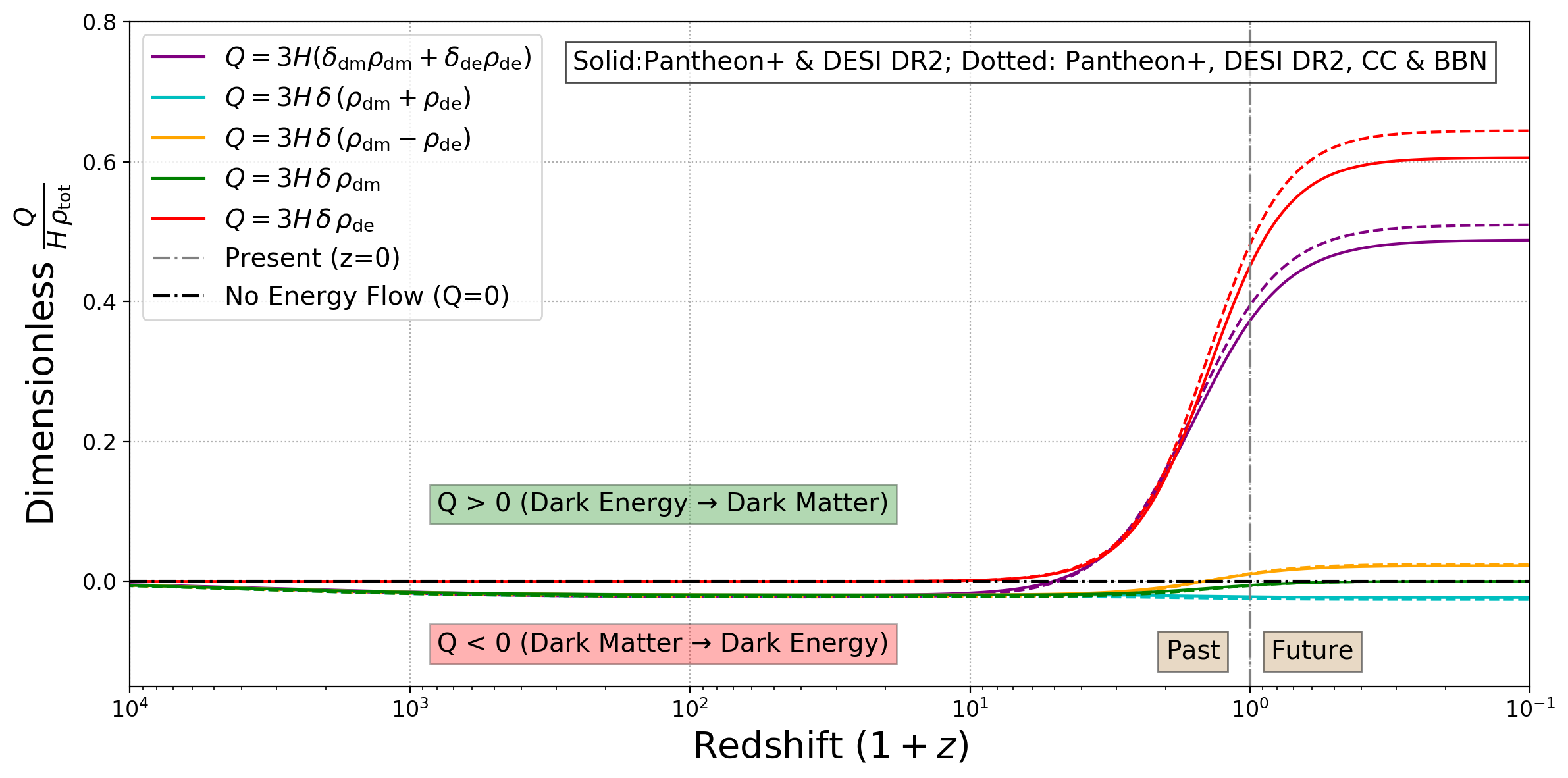}
        \caption{Direction of energy transfer vs redshift for linear interactions, using the mean values obtained in Tables~\ref{tab:constraints_pantheon_desi} and~\ref{tab:constraints_all_models}. The direction of energy transfer is also summarised in Table~\ref{tab:Consequenses}. Here we see that the most general form, $Q=3 H (\delta_{\text{dm}} \rho_{\text{dm}} + \delta_{\text{de}} \rho_{\text{de}})$, exhibits sign-switching behaviour. It behaves similarly to $Q=3 H \delta \rho_{\text{dm}}$ in the distant past, with a small relative energy flow from dark matter to dark energy, but behaves closer to $Q=3 H \delta \rho_{\text{de}}$ in the more recent past and the predicted future, where a larger relative energy flow from dark energy to dark matter is observed. This sign-changing behaviour, with similar initial and final directions of energy transfer, is also exhibited by $Q=3 H \delta (\rho_{\text{dm}}-\rho_{\text{de}})$. For $Q=3 H \delta (\rho_{\text{dm}}+\rho_{\text{de}})$, the coupling to dark matter is more prominent than the coupling to dark energy, causing the energy flow to resemble that of $Q=3 H \delta \rho_{\text{dm}}$. For all models, the relative interaction strength diminishes during radiation domination at very high redshift.}
        \label{fig:Q_direction_linear}
    \end{minipage}

    \vspace{1cm}

    \begin{minipage}{0.95\linewidth}
        \centering
        \includegraphics[width=0.9\linewidth]{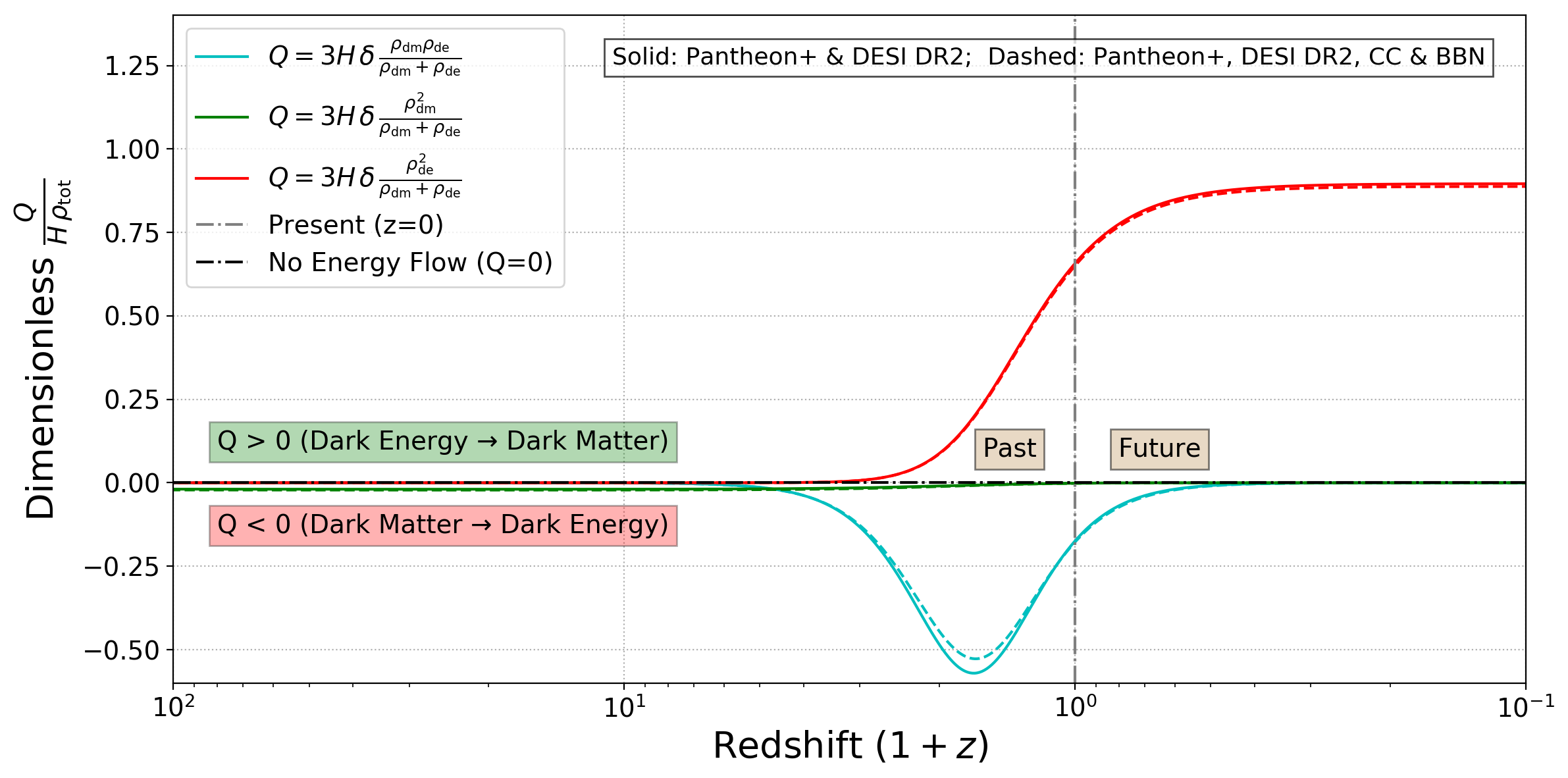}
        \caption{Direction of energy transfer vs redshift for non-linear interactions, using the mean values obtained in Tables~\ref{tab:constraints_pantheon_desi} and~\ref{tab:constraints_all_models}. The direction of energy transfer is also summarised in Table~\ref{tab:Consequenses}. For $Q=3H\delta \left( \frac{\rho_{\text{dm}}\rho_{\text{de}} }{\rho_{\text{dm}}+\rho_{\text{de}}} \right)$ and $Q=3H\delta \left( \frac{\rho^2_{\text{de}} }{\rho_{\text{dm}}+\rho_{\text{de}}} \right)$, the relative interaction strength diminishes at high redshift, when dark energy provides only a small contribution to the total energy density. Similarly, for $Q=3H\delta \left( \frac{\rho_{\text{dm}}\rho_{\text{de}} }{\rho_{\text{dm}}+\rho_{\text{de}}} \right)$ and $Q=3H\delta \left( \frac{\rho^2_{\text{dm}} }{\rho_{\text{dm}}+\rho_{\text{de}}} \right)$, which both involve energy transfer from dark matter to dark energy, the interaction becomes weaker in the future as the dark matter density becomes negligible.}
        \label{fig:Q_direction_nonlinear}
    \end{minipage}

\end{figure}

\begin{figure}
    \centering
    \includegraphics[width=0.99 \linewidth]{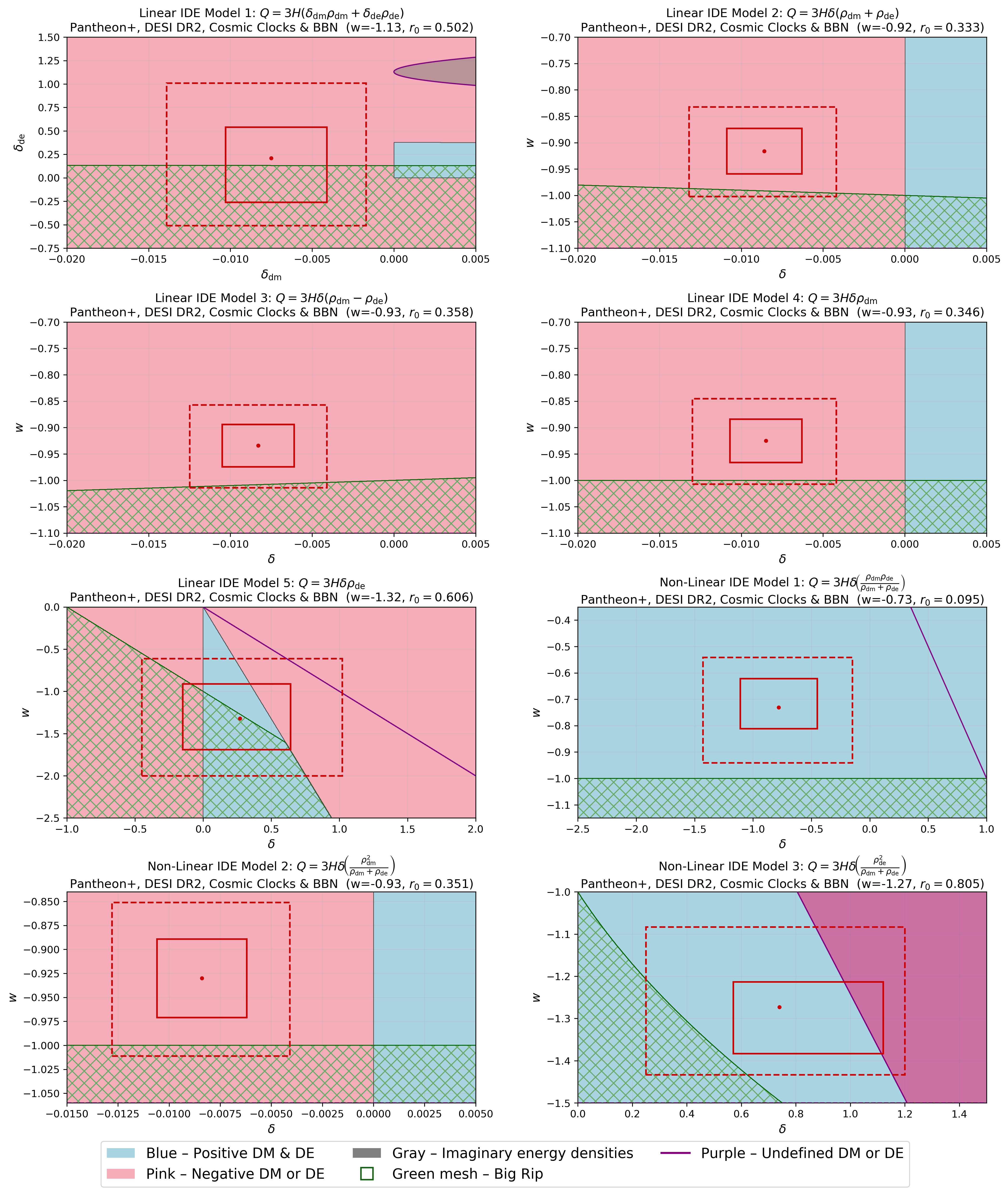}
    \caption{2D portraits showing the mean values obtained from Pantheon+, DESI DR2, Cosmic Clocks \& BBN data, as reported in Tables~\ref{tab:constraints_pantheon_desi} and~\ref{tab:constraints_all_models}, and how they correspond to the parameter space of each of the eight IDE models, using Tables~\ref{tab:Com_real} and~\ref{tab:Com_PEC}. The red solid, dashed, and dotted lines indicate the 68\% and 95\% confidence intervals. Blue areas indicate regions where the model has positive energy densities throughout the entire cosmic evolution. Pink areas indicate that negative energy densities occur either in the past or in the future. The gray overlay indicates the presence of imaginary energy densities, while the purple areas show undefined energy densities. Lastly, the green mesh indicates the presence of future big rip singularities. The purple areas in the bottom-right panel may have undefined values at some scale factor, but this is not guaranteed; see the conditions in Table~\ref{tab:Com_real}.}
    \label{fig:parameter_space_BBN}
\end{figure}

\begin{figure}
    \centering
    \includegraphics[width=0.7 \linewidth]{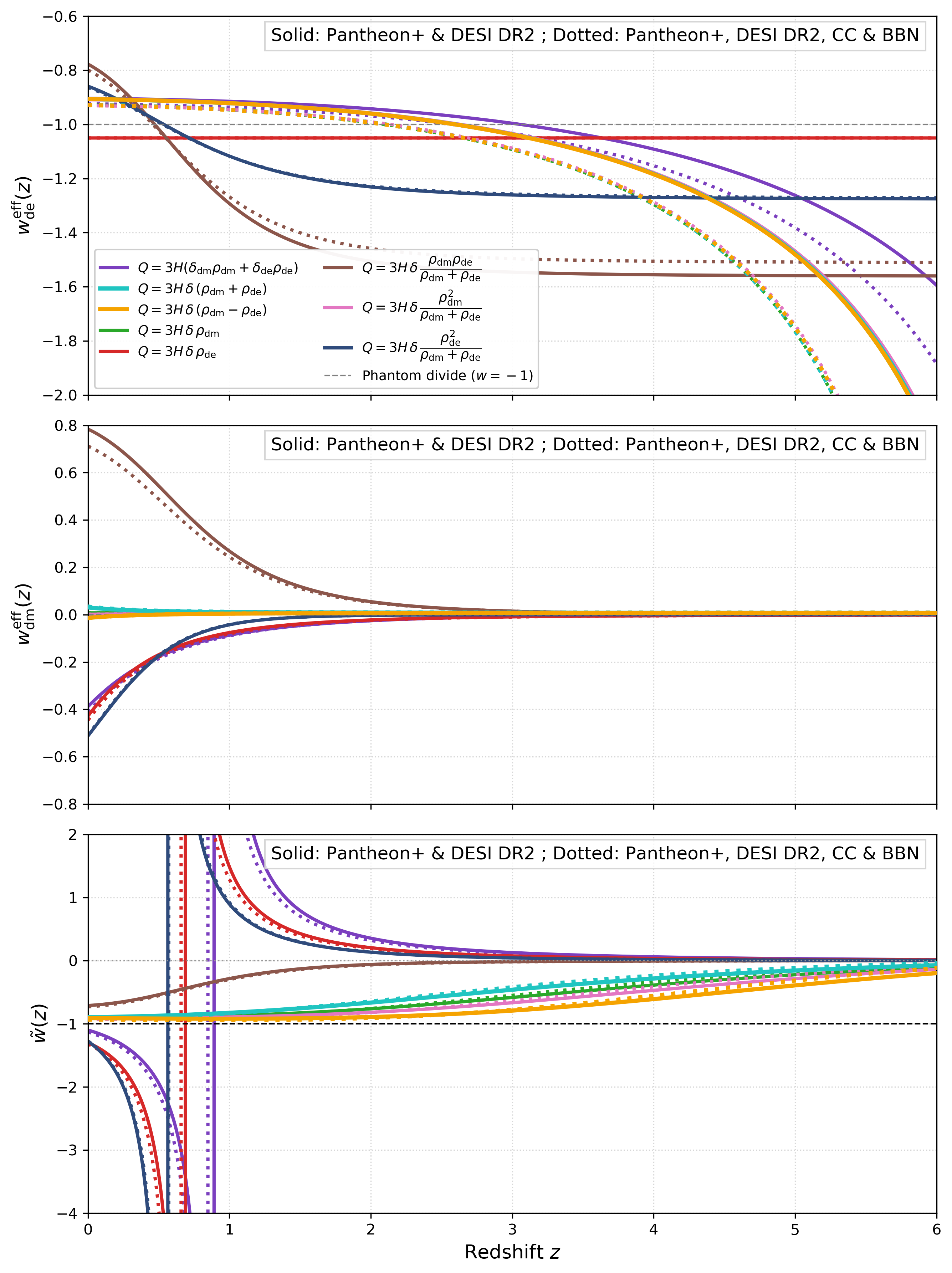}
    \caption{Evolution of the effective dark energy and dark matter equations of state, $w_{\rm{de}}^{\rm{eff}}(z)$ and $w_{\rm{dm}}^{\rm{eff}}(z)$, from \eqref{omega_eff_dm_de} (top and middle panels), and the reconstructed dark energy equation of state $\tilde{w}(z)$ from \eqref{wz_2} (bottom panel), using the mean values reported in Tables~\ref{tab:constraints_pantheon_desi} and~\ref{tab:constraints_all_models}. In the top panel, phantom-crossing behaviour, where the dark energy density decreases ($w_{\rm{de}}^{\rm{eff}}>-1$) at low redshift and increases ($w_{\rm{de}}^{\rm{eff}}<-1$) at higher redshift, is observed in seven of the eight models, with the only exception being $Q=3H\delta\rho_{\text{de}}$, which has a constant $w_{\rm{de}}^{\rm{eff}}$~\cite{vanderWesthuizen:2025vcb}. The curves corresponding to $Q=3H\delta(\rho_{\text{dm}}+\rho_{\text{de}})$, $Q=3H\delta(\rho_{\text{dm}}-\rho_{\text{de}})$, $Q=3H\delta\rho_{\text{dm}}$, and $Q=3H\delta\left(\frac{\rho_{\text{dm}}^2}{\rho_{\text{dm}}+\rho_{\text{de}}}\right)$ lie almost exactly on top of each other and are nearly indistinguishable. These four models, together with $Q=3H(\delta_{\text{dm}}\rho_{\text{dm}}+\delta_{\text{de}}\rho_{\text{de}})$, exhibit energy flow from dark matter to dark energy at high redshift and therefore undergo an additional divergent phantom crossing when the dark energy density becomes negative, as discussed in~\cite{vanderWesthuizen:2025vcb,vanderWesthuizen:2025mnw}. In the middle panel, the same four models remain nearly indistinguishable and stay close to $w_{\rm{dm}}^{\rm{eff}}\simeq0$, preserving cold dark matter behaviour. Models with $Q\propto\rho_{\rm{de}}$ and energy flow from dark energy to dark matter at low redshift, namely $Q=3H(\delta_{\text{dm}}\rho_{\text{dm}}+\delta_{\text{de}}\rho_{\text{de}})$, $Q=3H\delta\rho_{\text{de}}$, and $Q=3H\delta\left(\frac{\rho_{\text{de}}^2}{\rho_{\text{dm}}+\rho_{\text{de}}}\right)$, exhibit $w_{\rm{dm}}^{\rm{eff}}<0$. Conversely, when energy flows from dark matter to dark energy at low redshift, specifically for $Q=3H\delta\left(\frac{\rho_{\text{dm}}\rho_{\text{de}}}{\rho_{\text{dm}}+\rho_{\text{de}}}\right)$ and $Q=3H\delta(\rho_{\text{dm}}-\rho_{\text{de}})$, one finds $w_{\rm{dm}}^{\rm{eff}}>0$. In the bottom panel, divergent behaviour of $\tilde{w}(z)$ is observed when energy flows from dark energy to dark matter, as indicated in Figures~\ref{fig:Q_direction_linear} and~\ref{fig:Q_direction_nonlinear} and summarized in Table~\ref{tab:Consequenses}, in agreement with the discussion in~\cite{vanderWesthuizen:2025mnw}.}
    \label{fig:w_eff} 
\end{figure}

When comparing the eight interaction models, several interesting trends emerge, which we summarize according to the following properties (with figures and tables illustrating or clarifying these statements indicated in brackets):

\begin{itemize}
    \item \textbf{Direction of energy transfer} (Figures~\ref{fig:Q_direction_linear} and~\ref{fig:Q_direction_nonlinear}, and Table~\ref{tab:Consequenses}): For the most general linear interaction, $Q=3H(\delta_{\text{dm}}\rho_{\text{dm}}+\delta_{\text{de}}\rho_{\text{de}})$, as well as for $Q=3H\delta(\rho_{\text{dm}}-\rho_{\text{de}})$, a preference is found for a sign-switching interaction, with energy flowing from dark matter to dark energy at early times (corresponding to $\delta_{\rm dm}<0$) and from dark energy to dark matter at later times (corresponding to $\delta_{\rm de}>0$). This behaviour is in agreement with recent reconstructions of the interaction kernel $Q$ from data, where sign-switching behaviour has been suggested in~\cite{Li:2025ula, Guedezounme:2025wav}.
    For all other interactions, sign-switching behaviour is not possible. Instead, for the remaining models we observe a preference for energy flow from dark matter to dark energy when $Q\propto\rho_{\rm dm}$ in any form. The only exceptions, where energy flows from dark energy to dark matter, occur for interactions proportional to $\rho_{\rm de}$, specifically $Q=3H\delta\rho_{\text{de}}$ and $Q=3H\delta \left( \frac{\rho^2_{\text{de}}}{\rho_{\text{dm}}+\rho_{\text{de}}} \right)$, which, as noted in Section~\ref{sec:stats}, exhibit the weakest statistical preference among the interactions considered.
    \item \textbf{Presence of negative energies} (Figure~\ref{fig:parameter_space_BBN} and Table~\ref{tab:Consequenses}): As discussed above, most models prefer energy flow from dark matter to dark energy, which in turn leads to negative dark energy densities in the past~\cite{vanderWesthuizen:2025vcb}, as illustrated in Figure~\ref{fig:parameter_space_BBN}. Other observational indications hinting at $\rho_{\rm de}<0$ in the past can also be found in~\cite{BOSS:2014hwf, Escamilla:2023shf, Malekjani:2023ple}. Conversely, interactions for which $Q\propto\rho_{\rm de}$ exhibit energy flow from dark energy to dark matter and therefore avoid negative energy densities, while the interaction $Q=3H\delta \left( \frac{\rho_{\text{dm}} \rho_{\text{de}} }{\rho_{\text{dm}}+\rho_{\text{de}}} \right)$ always yields positive energy densities across the entire parameter space.
    \item \textbf{Future big rip singularities} (Figure~\ref{fig:parameter_space_BBN}): Future big rip singularities can also occur, and are most likely for the interaction $Q=3H\delta\rho_{\text{de}}$ (see the overlap between the green mesh and the contour plots in Figure~\ref{fig:parameter_space_BBN}), although the parameter space is not sufficiently well constrained to draw definitive conclusions. For the same interaction, it is worth noting that the 65\% and 95\% confidence intervals may cross the upper positive-energy limit for $\delta$, which could lead to a past non-singular bounce, a feature of interacting models that will be discussed in detail in future work.  
    We again emphasise that the predicted negative energies and future singularities do not necessarily represent pathologies of the theory, as they may instead indicate that the model has been pushed beyond its domain of applicability.
    \item \textbf{Possible early-time instabilities} (Figure~\ref{fig:parameter_space_BBN} and Table~\ref{tab:doom}): As discussed in Section~\ref{theo_const}, the early-time stability of the models may be assessed \emph{a priori} from the sign of the doom factor $\textbf{d}$ in \eqref{DSA.doom}, where $\textbf{d}<0$ implies a stable model. Using the mean values obtained in the previous section, this analysis leads to the results summarized in Table~\ref{tab:doom}, where half of the models are found to be free from instabilities. We emphasize once more that this analysis is preliminary, and that a full perturbation study for each model will be carried out in future work.
    \item \textbf{Phantom crossing of $w_{\rm de}^{\rm eff}(z)$} (Figure~\ref{fig:w_eff}): A clear trend is observed in the evolution of $w_{\rm de}^{\rm eff}(z)$ in the top panel of Figure~\ref{fig:w_eff}, which indicates when $\rho_{\rm de}$ increases or decreases over time. For seven of the eight models, we observe phantom-crossing behaviour, with the dark energy density decreasing at low redshift ($w_{\rm de}^{\rm eff}>-1$) and increasing at higher redshift ($w_{\rm de}^{\rm eff}<-1$). The only exception is $Q=3H\delta\rho_{\text{de}}$, for which $w_{\rm de}^{\rm eff}=w+\delta$ is constant~\cite{vanderWesthuizen:2025vcb}. This increase and subsequent decrease of dark energy density was also suggested by the DESI collaboration~\cite{DESI:2025fii}, and may be explained by interactions between the dark sectors, as noted in~\cite{Guedezounme:2025wav}.
    Care is required when comparing these results, since our models also involve dark matter with non-cold behaviour, $w_{\rm dm}^{\rm eff}\neq 0$, as illustrated in the middle panel of Figure~\ref{fig:w_eff}. Finally, as discussed in Section~\ref{phantom_background}, interaction models with energy flow from dark energy to dark matter also exhibit a divergent phantom crossing for the reconstructed dark energy equation of state $\tilde{w}(z)$ from \eqref{wz_2}. As shown in the bottom panel of Figure~\ref{fig:w_eff}, these models have $\tilde{w}(z)<-1$ at low redshift and $\tilde{w}(z)>-1$ at higher redshift. In contrast, interactions for which energy flows from dark matter to dark energy do not exhibit any phantom-crossing behaviour.    
\end{itemize}

\begin{table}[h!]
\centering
\renewcommand{\arraystretch}{1.1} % Adjust row spacing
\setlength{\tabcolsep}{10pt}     % Adjust column spacing
\begin{tabular}{|c|c|c|c|c|c|}
\hline
\textbf{Interaction $Q$} 
 & \text{Preferred energy flow} 
 & $\rho_{\rm{dm},past}$ 
 & $\rho_{\rm{dm},future}$   & $\rho_{\rm{de,past}}$
 & $\rho_{\rm{de},future}$
\\ \hline \hline

$3 H (\delta_{\text{dm}} \rho_{\text{dm}} + \delta_{\text{de}}  \rho_{\text{de}})$
 & \text{DM $\rightarrow$ DE followed by DE $\rightarrow$ DM}
 & + & + & -  & +
\\ \hline

$3H\delta( \rho_{\text{dm}}+\rho_{\text{de}})$
 & \text{DM $\rightarrow$ DE}
 & + & - & -  & +
\\ \hline

$3H\delta( \rho_{\text{dm}}-\rho_{\text{de}})$
 & \text{DM $\rightarrow$ DE followed by DE $\rightarrow$ DM}
 & + & + & -  & +
\\ \hline

$3H \delta \rho_{\text{dm}}$
  & \text{DM $\rightarrow$ DE}
 & + & + & -  & +
\\ \hline

$3H\delta \rho_{\text{de}}$
 & \text{DE $\rightarrow$ DM}
 & + & + & +  & +
\\ \hline

$3H\delta \left( \frac{\rho_{\text{dm}} \rho_{\text{de}} }{\rho_{\text{dm}}+\rho_{\text{de}}} \right)$
 & \text{DM $\rightarrow$ DE}
 & + & + & +  & +
\\ \hline

$3H\delta \left( \frac{\rho^2_{\text{dm}} }{\rho_{\text{dm}}+\rho_{\text{de}}} \right)$
  & \text{DM $\rightarrow$ DE}
 & + & + & -  & +
\\ \hline

$3H\delta \left( \frac{\rho^2_{\text{de}} }{\rho_{\text{dm}}+\rho_{\text{de}}} \right)$
 & \text{DE $\rightarrow$ DM}
 & + & + & +  & +
\\ \hline

\end{tabular}
\caption{Summary of the inferred direction of energy flow and the positivity of dark matter and dark energy for each model, based on the posteriors reported in Tables~\ref{tab:constraints_pantheon_desi} and~\ref{tab:constraints_all_models}. It should be noted that, since not all parameters are well constrained, alternative scenarios are still allowed. The direction of energy flow is visualized for linear models in Figure~\ref{fig:Q_direction_linear} and for non-linear models in Figure~\ref{fig:Q_direction_nonlinear}. See Table~\ref{tab:Com_PEC} and Figure~\ref{fig:parameter_space_BBN} for a more detailed understanding of the parameter space and its consequences for each model.}
\label{tab:Consequenses}
\end{table}

\begin{table}[h!]
\centering
\renewcommand{\arraystretch}{1.1} % Adjust row spacing
\setlength{\tabcolsep}{10pt}     % Adjust column spacing
\begin{tabular}{|c|c|c|c|c|}
\hline
\textbf{Interaction $Q$} 
 & \text{Sign of Q} 
 & \text{Sign of $(1+w)$}
  & \text{Sign of $\rho_{\rm{de}}$}
 & \text{A priori stable $(\textbf{d}<0)$} 
\\ \hline \hline

$3 H (\delta_{\text{dm}} \rho_{\text{dm}} + \delta_{\text{de}}  \rho_{\text{de}})$
& -
 & -
  & -
 & $\checkmark$ 
\\ \hline

$3H\delta( \rho_{\text{dm}}+\rho_{\text{de}})$
& -
 & +
  & -
 & $X$  \\ \hline

$3H\delta( \rho_{\text{dm}}-\rho_{\text{de}})$
& - 
 & +
  & -
 & $X$  
\\ \hline

$3H \delta \rho_{\text{dm}}$
& - 
 & +
  & -
 & $X$  
\\ \hline

$3H\delta \rho_{\text{de}}$
& + 
 & -
  & +
 & $\checkmark$ 
\\ \hline

$3H\delta \left( \frac{\rho_{\text{dm}} \rho_{\text{de}} }{\rho_{\text{dm}}+\rho_{\text{de}}} \right)$
& - 
 & +
  & +
 & $\checkmark$ 
\\ \hline

$3H\delta \left( \frac{\rho^2_{\text{dm}} }{\rho_{\text{dm}}+\rho_{\text{de}}} \right)$
& - 
 & +
  & -
 & $X$  
\\ \hline

$3H\delta \left( \frac{\rho^2_{\text{de}} }{\rho_{\text{dm}}+\rho_{\text{de}}} \right)$
& + 
 & +
  & +
 & $X$
\\ \hline

\end{tabular}
\caption{Summary of the inferred stability for each interaction from the doom-factor analysis, using the sign of $\textbf{d}$ in equation~\eqref{DSA.doom}, based on the posteriors obtained in Tables~\ref{tab:constraints_pantheon_desi} and~\ref{tab:constraints_all_models}. The sign of each quantity is evaluated at early times, or equivalently at high redshift. It should be noted that, since not all parameters are well constrained, alternative possibilities are still allowed.}

\label{tab:doom}
\end{table}

\section{Conclusions}\label{sec:conclusions}

In this study, we constrained interacting dark energy models using two different combinations of late-time datasets. The first combination included only Pantheon+ Type Ia supernovae and DESI DR2 baryon acoustic oscillation data, while the second combination additionally incorporated Cosmic Clocks and Big Bang Nucleosynthesis.
The models considered belong to a class of phenomenological interacting dark energy models for which analytical solutions for the dark matter and dark energy densities, and consequently for the Hubble parameter $H(z)$, have been derived in~\cite{vanderWesthuizen:2025vcb, vanderWesthuizen:2025mnw}. These include five linear IDE models, namely the most general case $Q=3H(\delta_{\rm dm}\rho_{\rm dm}+\delta_{\rm de}\rho_{\rm de})$ and four special cases: $Q=3H\delta(\rho_{\rm dm}+\rho_{\rm de})$, $Q=3H\delta(\rho_{\rm dm}-\rho_{\rm de})$, $Q=3H\delta\rho_{\rm dm}$, and $Q=3H\delta\rho_{\rm de}$. In addition, we studied three non-linear IDE models: $Q=3H\delta\left(\tfrac{\rho_{\rm dm}\rho_{\rm de}}{\rho_{\rm dm}+\rho_{\rm de}}\right)$, $Q=3H\delta\left(\tfrac{\rho_{\rm dm}^2}{\rho_{\rm dm}+\rho_{\rm de}}\right)$, and $Q=3H\delta\left(\tfrac{\rho_{\rm de}^2}{\rho_{\rm dm}+\rho_{\rm de}}\right)$.

For all eight IDE models, we find a better fit than $\Lambda$CDM from a $\Delta\chi^2$ analysis for both combinations of datasets considered, as shown in Table~\ref{tab:model_statistics}. When using the Akaike Information Criterion ($\Delta$AIC), we find a similarly improved fit in all cases except for $Q=3H\delta\rho_{\rm de}$ when Cosmic Clocks and BBN data are added. In all cases, the improvement is largest when using only Pantheon+ and DESI DR2 data, and slightly diminishes when Cosmic Clocks and BBN are included.
We also find that models in which $Q\propto\rho_{\rm dm}$ in any form exhibit the strongest statistical preference, and are accompanied by tighter constraints on $\delta_{\rm dm}$ or $\delta$. In contrast, models with $Q\propto\rho_{\rm de}$, specifically $Q=3H\delta\rho_{\rm de}$ and $Q=3H\delta\left(\tfrac{\rho_{\rm de}^2}{\rho_{\rm dm}+\rho_{\rm de}}\right)$, show a weaker preference and significantly larger uncertainties on $\delta_{\rm de}$ or $\delta$.

In terms of the direction of energy transfer, our analysis shows a preference for sign-switching interactions, with energy transfer from dark energy to dark matter at low redshift and from dark matter to dark energy at higher redshift, in agreement with recent reconstructions~\cite{Li:2025ula, Guedezounme:2025wav}. Models that do not allow sign-changing behaviour instead show a preference for energy flow from dark matter to dark energy for all cases in which $Q\propto\rho_{\rm dm}$ in any form, with the exceptions of $Q=3H\delta\rho_{\rm de}$ and $Q=3H\delta\left(\tfrac{\rho_{\rm de}^2}{\rho_{\rm dm}+\rho_{\rm de}}\right)$, which instead exhibit energy flow from dark energy to dark matter. These results are illustrated in Figures~\ref{fig:Q_direction_linear} and~\ref{fig:Q_direction_nonlinear}, and summarized in Table~\ref{tab:Consequenses}.
Caution is required when interpreting these findings, since all models that exhibit energy transfer from dark matter to dark energy, except $Q=3H\delta\left(\tfrac{\rho_{\rm dm}\rho_{\rm de}}{\rho_{\rm dm}+\rho_{\rm de}}\right)$, which always produces positive energy densities, are accompanied by negative dark energy densities in the past, which may be physically problematic (see Figure~\ref{fig:parameter_space_BBN} and Table~\ref{tab:Consequenses}). It may nevertheless be argued that negative dark energy does not necessarily constitute a pathology, since the total dark sector remains conserved and positive. In this sense, a unified dark fluid (UDF) description of interacting dark energy models may reproduce the same expansion history while avoiding the apparent pathologies of a two-fluid description. This possibility deserves further investigation and will be explored in future work; connections between UDF and IDE models can already be found in~\cite{Zhang:2004gc, Li:2014eha, vonMarttens:2019ixw, Kou:2025yfr, Aguilar-Perez:2025iqp}.
Alternatively, negative energy densities may simply indicate that the phenomenological models considered here are being extrapolated beyond their domain of applicability.

Furthermore, for all interactions considered, we find a phantom-divide crossing in the effective dark energy equation of state $w^{\rm{eff}}_{\rm{de}}(z)$, with dark energy decreasing ($w^{\rm{eff}}_{\rm{de}}>-1$) at present and at low redshift, and increasing ($w^{\rm{eff}}_{\rm{de}}<-1$) at higher redshift in the more distant past. This behaviour is illustrated, together with the evolution of $w^{\rm{eff}}_{\rm{dm}}(z)$ and $\tilde{w}(z)$, in Figure~\ref{fig:w_eff}. Divergent behaviour of $w^{\rm{eff}}_{\rm{de}}(z)$ is not shown in the figure, but is also expected at higher redshift when $\rho_{\rm{de}}=0$, before crossing into negative values.
It should be kept in mind that these results are illustrative rather than definitive, since the plots are constructed using only the mean values of the posterior distributions. Different conclusions may therefore be drawn when exploring other regions of the parameter space allowed by the uncertainties. 

A quick comparison of the eight models in Tables~\ref{tab:Consequenses} and~\ref{tab:doom} shows that $Q=3H\delta\left(\tfrac{\rho_{\text{dm}}\rho_{\text{de}}}{\rho_{\text{dm}}+\rho_{\text{de}}}\right)$ and $Q=3H\delta\rho_{\rm de}$ are the least problematic cases, as both models avoid negative energy densities and early-time instabilities. There is an additional theoretical preference for $Q=3H\delta\rho_{\rm de}$, since energy transfer from dark energy to dark matter is favoured by thermodynamic arguments~\cite{Pavon:2007gt}. However, this model also exhibits the weakest statistical performance, as shown in Table~\ref{tab:model_statistics}.

In conclusion, our analysis shows that interacting dark energy models are promising but also challenging. While these models can fit the data well and provide a possible mechanism for dynamical dark energy, an increasingly topical subject, they are accompanied by a complex parameter space. In several cases, alleviating existing tensions comes at the expense of introducing more severe issues, such as negative energy densities and early-time instabilities. We therefore believe that these models warrant further investigation, extending the analysis to include perturbations, early-time datasets, and interaction models that go beyond purely phenomenological descriptions toward a more robust microphysical foundation.

\begin{acknowledgments}

EDV is supported by a Royal Society Dorothy Hodgkin Research Fellowship. This article is based upon work from the COST Action CA21136 - ``Addressing observational tensions in cosmology with systematics and fundamental physics (CosmoVerse)'', supported by COST - ``European Cooperation in Science and Technology''.
\end{acknowledgments}

\bibliography{bib_paper}  

\end{document}